\documentclass{ws-ijmpe}
\usepackage[super,compress]{cite}
\usepackage{subcaption}
\newcommand*\atom[3]{\ensuremath{{}^{#3}_{#2}\mathrm{#1}}}
\begin{document}

\markboth{Authors' Names}{Instructions for typing manuscripts (paper's title)}

\catchline{}{}{}{}{}

\title{Effects of transient non-thermal particles on the big bang nucleosynthesis}

\author{Tae-Sun Park$^{\dagger,\dagger\dagger}$, Kyung Joo Min$^{\ddagger}$ \and Seung-Woo Hong$^{\dagger,\ddagger,*}$}

\address{$^{\dagger}$Department of Physics, Sungkyunkwan University\\
Suwon 16419, Korea\\
$^{\dagger\dagger}$Center for Exotic Nuclei Studies, Institute for Basic Science\\
Daejeon 34126, Korea\\
$^{\ddagger}$Department of Energy Science, Sungkyunkwan University\\
Suwon 16419, Korea\\
\footnote{Corresponding author} swhong@skku.ac.kr}

\maketitle

\begin{history}
\received{Day Month Year}
\revised{Day Month Year}
\end{history}

\begin{abstract}
The effects of introducing a small amount of
non-thermal distribution (NTD) of elements in big bang nucleosynthesis (BBN)
are studied by
allowing a fraction of the NTD
to be time-dependent so that it contributes
only during a certain period of the BBN evolution.
The fraction is
modeled as a Gaussian-shaped function
of $\log(T)$, where
$T$ is the temperature of the cosmos,
and thus the function is specified by three parameters;
the central temporal position, the width and the magnitude.
The change in the average nuclear reaction rates due to the presence of the NTD
is assumed to be proportional to the
Maxwellian reaction rates
but with temperature $T_{\rm NTD} \equiv \zeta T$,
$\zeta$ being another parameter of our model.
By scanning a wide four-dimensional parametric space
at about half a million points,
we have found
about 130 points with $\chi^2< 1$,
at which the predicted primordial abundances of light elements
are consistent with the observations.
The magnitude parameter $\varepsilon_0$
of these points turns out to be
scattered over a very wide range
from $\varepsilon_0 \sim 10^{-19}$
to $\sim 10^{-1}$,
and
the $\zeta$-parameter is found to be
strongly correlated with the magnitude parameter $\varepsilon_0$.
The temperature region with $0.3\times 10^9 \mbox{K} \lesssim T \lesssim 0.4\times 10^9 \mbox{K}$
or
the temporal region $t\simeq 10^3$ s
seems to play a central role
in lowering $\chi^2$.
\end{abstract}


\ccode{PACS numbers: 26.35.+c, 26.40.+r, 98.80.Ft}


\section{Introduction}

Big bang nucleosynthesis (BBN)
serves as the standard scenario
to address the primordial abundances of light elements
of our universe \cite{alpher,schramm,peebles,smith,sarkar,steigman}.
With the baryon-to-photon ratio $\eta_0$ determined accurately
by the cosmic microwave background radiation
measurement of
Wilkinson Microwave Anisotropy Probe~\cite{spergel}
and Planck~\cite{planck},
BBN is basically parameter-free
and can explain the abundances of primordial
deuteron and $^4$He quite successfully.
However, the BBN prediction for lithium is
reported to be
about three times bigger than the 
observation \cite{melendez,asplund,kusakabe,cyburt,iocco},
and has brought a lot of attention 
(see Refs.~\citen{fields,israelian}, for instance).

Incidentally,
a puzzling drop in Li/H 
in metal-poor stars has been observed\cite{sbordone},
which imposes a substantial uncertainty on the primordial Li abundance.
We refer to, for example,
Refs.~\citen{korn,fu,spite,ioccob}
for stellar models
that take into account detailed mechanisms of
the Li depletion in those stars.
These models, however,
do not resolve the Li discrepancy completely~\cite{PDG2018}.


There have been theoretical efforts to resolve the above
so-called ``lithium problem"
by altering
the assumption that
all particles except neutrinos are at thermal equilibrium with
the Maxwellian distribution.
For example,
the in-flight reaction probability
has been extensively studied \cite{voroncheva,voronchevb}
by taking into account the fact that
the particles created from nuclear reactions
can have energies in the MeV range
and thus can have a probability of overcoming the Coulomb repulsion
to go through nuclear reactions
before being thermalized.
The effect of such a mechanism turns out to be
insufficient to change the abundance significantly.
Another interesting approach was attempted by Bertulani et al.\cite{bertulani}, 
where the Maxwell-Boltzmann(MB) distribution was modified to adopt 
the so-called non-extensive statistics
with the results that it only worsens the lithium problem.

Our study is closely related to the work by Kang et al.\cite{kang}, 
where they have introduced so-called ``cosmic rays",
which consist of
only proton isotopes (protons, deuterons and tritons)
that are assumed to have a power-law shaped distribution up to 4 MeV.
In the study of Kang et al.\cite{kang},
the upper limit of the distribution is severely constrained
since cosmic rays with energies higher than the
D($p$, $n$)$\mbox{H}_{2}$ threshold, 3.337 MeV, destroy deuterium.
By tuning the fraction of the cosmic rays 
with respect to thermal isotopes of hydrogen to 
$0.7\times 10^{-6}$,
they could account for the lithium abundance successfully,
but with a 5 \% reduction of the deuteron abundance.
In their approach, 
the fraction of particles
(denoted by $\varepsilon$)
with non-Maxwellian distribution
was treated as time-independent.
This assumption may be questioned 
in that
the cosmic rays with such a non-MB distribution may exist only 
for a certain period 
rather than being independent of time or temperature.

The question on
the origin of the time-dependent cosmic rays
is not addressed here,
but there are candidates related to, but not limited to,
decay or annihilation of relic particles,
which would inject cosmic rays to the universe.
The outgoing particles of decay processes
of relic particles
would be rapidly thermalized due to the interactions
with background materials,
but the decay or annihilation rate will have a non-trivial time-dependence
characterized by the
lifetimes or the time-dependence of the energies and densities of the relic particles.
For example, 
the decay of
the next lightest supersymmetric particle (NLSP)
into the 
lightest supersymmetric particle (LSP) dark matter
can produce time-dependent suprathermal particles.
Indeed, it was discussed that
the stau-NLSP and gravitino-LSP system
with stau lifetime $\tau \simeq 10^3\ \mbox{s}$ 
could resolve the lithium problem with some representative values of
the model parameters\cite{bailly}.
We refer to Ref. \refcite{jedamzik} and references therein for a comprehensive
review on the role of dark matter on the BBN.
%
%

In this work, we do not stick to any particular scenario or candidate 
for the origin of cosmic rays.
%
Instead, we explore 
the possibility in which
the fraction of the NTD is
a function of time or temperature.
Since temperature is a monotonically
decreasing function of time,
time dependence can be converted to temperature dependence.
The time dependence of the fraction $\varepsilon(T)$ is modeled as a
Gaussian-shaped function of $\log(T)$ with three parameters for the
central temporal position $T_0$, the width $\Delta$, 
and the magnitude $\varepsilon_0$:
An explicit functional form will be given in the next section.
Although a power-law type distribution would be a reasonable approach,
making a realistic model for 
the non-MB distribution
requires an initial condition and the evolution with time.
Thus,
we make a naive assumption
for computational convenience
by assuming that
the averaged reaction rate
$\langle \sigma v \rangle_{ij\to kl} (T)$
in the presence of the NTD
can be approximated as a sum of two
Maxwellian reaction rates of particles at temperatures $T$ 
and $T_{\rm NTD}$,
{\it i.e.},
\begin{equation}
\langle \sigma v \rangle_{ij\to kl} (T)
= \left[1-\varepsilon(T)\right] R_{ij\to kl}^{th}(T) 
+ \varepsilon(T) R_{ij\to kl}^{th}(T_{NTD}),
\label{assume}
\end{equation}
where the subscript ``$ij\to kl$" is the reaction index,
$T_{\rm NTD}$ stands for ``the temperature of the NTD",
and
$R_{ij\to kl}^{th}(T)$ is the usual Maxwell-Boltzmann averaged
reaction rate in thermal equilibrium at temperature $T$.
We introduce $\zeta \equiv T_{\rm NTD}/T$ and treat $\zeta$ as a free parameter.
Since the non-thermal particles with temperatures less than $T$ 
would not change the reaction rates significantly,
the parameter $\zeta$
is expected to be larger than unity. 
This assumption of a Maxwell-Boltzmann reaction rate for the NTD
simplifies the calculation significantly.

We then scan a wide range of 
four-dimensional parameter space of $(\varepsilon_0,\,T_0,\, \Delta,\, \zeta)$,
searching for the best parameters
that meet the observational data for the primordial abundances.
The details of our model and the calculational method 
are described in Sec. II,
which is followed by the 
results in Sec. III,
and discussions in Sec. IV.

\section{Calculational method}

Let us begin with the discussion of the assumption in Eq.~(\ref{assume}).
The averaged reaction rate of the reaction 
$i+j \to k + l$
reads
\begin{eqnarray}
\langle\sigma v\rangle_{ij\to kl}(T)
&=&
\frac{1}{2} \int_{-1}^1\! d\cos \theta_{ij}
\int_0^\infty\!\! dE_i\int_0^\infty\!\! dE_j
\nonumber \\
&&
\ \ \
\, v_{ij}\, \sigma_{ij\to kl}
\, f_i(E_i,T)f_j(E_j,T),
\label{sigmavij}
\end{eqnarray}
where 
$\theta_{ij}$ is the relative angle,
$v_{ij}$ is the relative velocity, 
$\sigma_{ij\to kl}$ is the
cross section of the reaction,
and $f_i(E_i,T)$ is the distribution of the $i$-th particle 
with energy $E_i$.
Equation~(\ref{sigmavij}) can be rewritten as an integration over 
the center-of-mass energy $E_{ij}$,
\begin{equation}
\langle\sigma v\rangle_{ij\to kl}(T)
= \int_0^\infty\! d E_{ij} \, v_{ij}\, \sigma_{ij\to kl}
\, F_{ij}(E_{ij},T).
\label{Rij0}
\end{equation}
When all the incoming particles are in thermal equilibrium
with Maxwellian distribution,
$f_i(E_i,\,T) = f^{th}(E_i,\, T) 
\equiv
2 \sqrt{ E_i / \pi (k T)^3} e^{-E_i /k T}$,
$F_{ij}(E_{ij}, T)$ becomes identical to the Maxwell-Boltzmann distribution,
$F_{ij}(E_{ij}, T) = f^{th}(E_{ij}, \, T)$.
In the presence of non-thermal components,
the $i$-th particle's distribution can be written as
\begin{equation}
f_i(E_i,T) = \left[1 - \varepsilon_i(T)\right] f^{th}(E_i, T)
 + \varepsilon_i(T) f_i^{ntd}(E_i, T)
\end{equation}
with the normalization condition
$\int_0^\infty d E_i f_i^{ntd}(E_i, T) = 1$,
where $f_i^{ntd} (E_i,T)$ is a yet unknown 
distribution of the non-thermal part,
and $\varepsilon_i(T)$ is its temperature-dependent magnitude.
Then $F_{ij}$ reads
in non-relativistic limit
\begin{eqnarray}
F_{ij}(E_{ij}, T) &=& 
(1-\varepsilon_i) (1- \varepsilon_j) f^{th}(E_{ij},\, T)
\nonumber \\
&&+\>
\int_0^\infty \!\! \!dE_i 
\int_{E_{j-}}^{E_{j+}}\!\!\! dE_j
\, 
\frac{{\cal F}_{ij}^{ntd}}{2 \mu v_i v_j} 
\end{eqnarray}
with
\begin{eqnarray}
{\cal F}_{ij}^{ntd} &=&
\varepsilon_i (1-\varepsilon_j)  f_i^{ntd}(E_i, T)f^{th}(E_j, T)
\nonumber \\
&&+\ (1-\varepsilon_i) \varepsilon_j  f^{th}(E_i, T)f_j^{ntd}(E_j,T)
\nonumber \\
&&+\ 
\varepsilon_i \varepsilon_j  f_i^{ntd}(E_i,T)f_j^{ntd}(E_j,T),
\label{Fij-full}
\end{eqnarray}
where
$\varepsilon_i=\varepsilon_i(T)$,
$\mu =m_i m_j/(m_i+m_j)$,
$v_{ij} = \sqrt{2E_{ij}/\mu}$,
$v_i = \sqrt{2 E_i/m_i}$,
and
$E_{j \pm} = \frac12 m_j (v_{ij}\pm v_i)^2$.
Instead of modeling $f_i^{ntd}$,
we assume that
$F_{ij}(E_{ij},T)$ 
may be effectively approximated as
\begin{equation}
F_{ij}(E_{ij}, T) = 
\left[1-\varepsilon\right(T)] f^{th}(E_{ij},\, T)
 + \varepsilon(T) f^{th}(E_{ij},\, T_{\rm NTD}),
\label{Fij0}
\end{equation}
where $\varepsilon(T)$ is the amount of the NTD
to be taken as $T$-dependent and is to be discussed shortly.
Combining Eqs.(\ref{Rij0}) and (\ref{Fij0}), 
and defining $R_{ij\to kl}^{th}(T)$ by
\begin{equation}
R_{ij\to kl}^{th}(T) \equiv
\int_0^\infty\! d E_{ij} \, v_{ij}\,\sigma_{ij\to kl}
\, f^{th}(E_{ij},\, T),
\label{Rdef}
\end{equation}
we are led to Eq.~(\ref{assume}).
One of the immediate advantages in this approach is that
we can make use of the well-established codes 
\cite{kawanoa,kawanob,wagonera,wagonerb,wagonerc,fowlera,fowlerb,harris,beaudet} 
available for the standard BBN calculations
without having to write extra codes
for the averaged reaction rates in the presence of NTD particles.

Before going further, let us remark on the reverse rates.
In the presence of NTD, our assumption given in Eq.~(\ref{assume})
implies the inverse rate $kl\to ij$ to be
\begin{equation}
\langle \sigma v \rangle_{kl\to ij} (T)
= \left[1-\varepsilon(T)\right] R_{kl\to ij}^{th}(T) 
+ \varepsilon(T) R_{kl\to ij}^{th}(T_{NTD}).
\label{inverse}
\end{equation}
As is the case with $R_{kl\to ij}^{th}(T)$,
the reverse rate $R_{kl\to ij}^{th}(T_{NTD})$ is computed
from $R_{ij\to kl}^{th}(T_{NTD})$
by using the time reversal symmetry
with the assumption of thermal equilibrium at temperature $T_{NTD}$.
Because both
$R_{ij\to kl}^{th}(T)$ and $R_{ij\to kl}^{th}(T_{NTD})$ satisfy the principle of the
detailed balance,
the total reverse rate also does.

\begin{figure}
\centering
\includegraphics[width=0.8\textwidth]{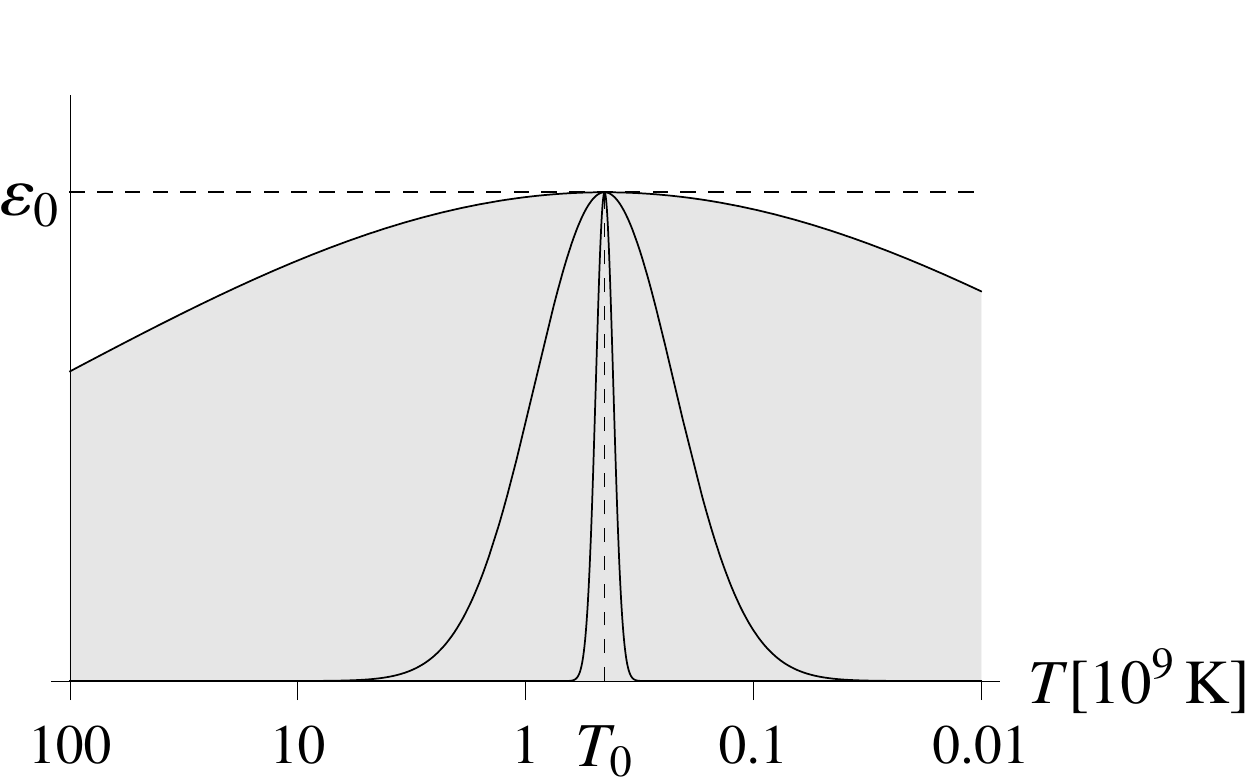}
\caption{The NTD portion $\varepsilon(T)$ for $T_0 = 0.45\times 10^9 \mbox{K}$, 
and $\Delta= 4$ (outermost), 1 (middle) and 0.0625 (innermost).}
\label{window-fig}
\end{figure}

In the consideration of 
the time or temperature dependence of the amount of NTD portion,
the dependence of $\varepsilon(T)$ on $T$ is taken as a
Gaussian-shaped function peaked at 
$T_0$ with a "window" of width $\Delta$ and peak height $\varepsilon_0$,
\begin{equation}
\varepsilon(T) = \varepsilon_0\, \exp\left[ 
- \left(\frac{\log(T/T_0)}{\Delta}\right)^2\right].
\label{window}
\end{equation}
Note that $\varepsilon(T)/\varepsilon_0 \ge 1/e$ only in the region
$T_0\, e^{-\Delta} \le T \le T_0\, e^{\Delta}$.
To illustrate the dependence of $\varepsilon(T)$ on $\Delta$, 
which corresponds to the length of time during which NTD particles appear,
we show $\varepsilon(T)$
for $T_0 = 0.45\times 10^9 \mbox{K}$ and three values of $\Delta$ in
Fig.~\ref{window-fig}.
Since the NTD portion is concentrated around $T\simeq T_0$,
$T_{\rm NTD}= \zeta T$ may be regarded as $T_{\rm NTD} \simeq \zeta T_0$,
especially when the width parameter $\Delta$ is small.

Here we would like to mention that
our choice for the shape of the fraction $\varepsilon(T)$
as a Gaussian 
is a naive ansatz chosen for computational convenience.
An exponentially decaying shape, for example, could be 
a more suitable choice for many scenarios of the NTD.
However,
since we will scan the parametric space with varying the width
of $\varepsilon(T)$,
the major effect due to the presence
of the NTD particle 
may be captured regardless of the details of the functional form.

Let us now describe how the parameter space is scanned.
For the magnitude parameter $\varepsilon_0$, we scan quite a large range
by choosing $n=1$ to $30$ in $\varepsilon_0 = 10^{-n}$
to see the possibility that even a very
tiny fraction of NTD may affect the lithium
problem. 
Most important nuclear synthesis occurs 
around $T \sim (0.1\sim 1)\times 10^9 \mbox{K}$, 
and thus for the peak temperature parameter $T_0$ 
we choose 41 grid points in the $\log(T_{0})$ axis from 
$T_{09} \equiv T_0/10^9 \mbox{K}  = 0.1$ to $1$  
by setting $T_{09} = 10^{n/40}$ with $n = -40$ to 0.
The width parameter $\Delta$ is chosen to vary
from $0.0625$ to $4$ by doubling the values of $\Delta$,
that is, $\Delta=2^n$ with $n=-4$ to 2.
As shown in Fig.~\ref{window-fig},
$\Delta=0.0625$ is narrow enough to 
explore the possibility of locating the period of time 
for non-thermal disturbance 
which can affect the BBN results,
and $\Delta=4$ is big enough for $\varepsilon(T)$
to be essentially regarded as independent of $T$.
The ratio of the NTD-temperature
to the temperature of the cosmos, $\zeta = T_{\rm NTD}/T$,
is chosen to be $\zeta = 10^{n/40}$ 
with $n\ge 1$.
Since the Kawano code is supposed to be accurate
only up to $T_9 \equiv T/10^9 \mbox{K} \lesssim 10$
and the reliability of the calculation for
$\zeta T_{09} \gtrsim 10$ is highly questionable,
we set an upper limit to
the $\zeta$ parameter
by imposing the condition
$\zeta T_{09} \le 10$,
i.e., $n\le 40$ for $T_{09}=1$ and $n \le 80$ for $T_{09}=0.1$.
In this search scheme, the total number of grid points considered is
$30 \times 41 \times 7 \times (40+80)/2 = 516,600$,
which can be summarized as follows:
\begin{eqnarray}
\varepsilon_0 &=& \left[ 10^{-1},\ 10^{-2},\ 10^{-3},\ \cdots,\ 10^{-30}\right],
\nonumber \\
T_{09} &=& \left[10^{-1}, 10^{-0.975}, 10^{-0.95}, \cdots,
\ 10^{0}\right],
\nonumber \\
\Delta &=& \left[ 0.0625,\ 0.125,\ 0.25,\  \cdots,\ 4\right],
\nonumber \\
\zeta &=& \left[10^{0.025},\ 10^{0.05},\ 10^{0.075},\ \cdots, 
10 /T_{09}
\right].
\end{eqnarray}

At each grid point, we evaluate $\chi^2$ defined by
\begin{equation}
\chi^2 = 
\left|\chi(\left.\mbox{D}/\mbox{H}\right|_p)\right|^2
+ \left|\chi(Y_{\rm p})\right|^2
+ \left|\chi(\left.\mbox{Li}/\mbox{H}\right|_p)\right|^2
\label{chi2-def}
\end{equation}
with
\begin{equation}
\chi(\alpha) \equiv \frac{{\cal C}(\alpha) - {\cal O}(\alpha)}{\sigma(\alpha)},
\label{chi-alpha}
\end{equation}
where ${\cal C}(\alpha)$, ${\cal O}(\alpha)$ and $\sigma(\alpha)$
stand for the calculated value, the observed value,
and the uncertainty 
for the quantity $\alpha$, respectively.
For ${\cal O}(\alpha)$ and $\sigma(\alpha)$, we adopt the PDG(2014) data
listed in Table~\ref{PDG-table}.

\begin{table*}
\centering
\caption{Particle Data Group data and the SBBN prediction for the primordial abundances.}
\label{PDG-table}
\begin{tabular}{lllll}
\hline
 & $\mbox{D}/\mbox{H}\ [10^{-5}]$  & $Y_{\rm p}$  &  $\mbox{Li}/\mbox{H}\ [10^{-10}]$ & Reference\\ 
\hline
PDG (2012) & $2.82 \pm 0.21$ & $0.249 \pm 0.009$ & $1.7 \pm 0.06 \pm 0.44$ & \citen{beringer}\\
PDG (2014) & $2.53 \pm 0.04$ & $0.2465 \pm 0.0097$ & $1.6 \pm 0.3$ & \citen{olive}\\
SBBN & $2.49\pm 0.17$ & $0.2486\pm 0.0002$ & $5.24^{ + 0.71}_{ - 0.62}$ & \citen{cyburt16}\\
\hline
\end{tabular}
\end{table*}

Here and hereafter, 
what we mean by $\mbox{Li}$ in Eq.~(\ref{chi2-def})
is the sum of
$^6\mbox{Li}$, $^7\mbox{Li}$ and $^7\mbox{Be}$.
This is because all the primordial 
$^7\mbox{Be}$ decays to $^7\mbox{Li}$,
and what is measured is the sum of
$^6\mbox{Li}$ and $^7\mbox{Li}$, not just $^7\mbox{Li}$, 
though the amount of primordial $^6\mbox{Li}$ is 
orders of magnitude less than that of $^7\mbox{Li}$.

For the calculation, we adopt the so-called 
Kawano code \cite{wagonera,fowlera,fowlerb,wagonerb,wagonerc,beaudet,harris,kawanoa,kawanob}.

\section{Results}
\subsection{Distribution of $\chi^2$ in the parameter space}

\begin{figure*}
\centering
%
\begin{subfigure}{.45\textwidth} \centering \includegraphics[width=0.95\textwidth]{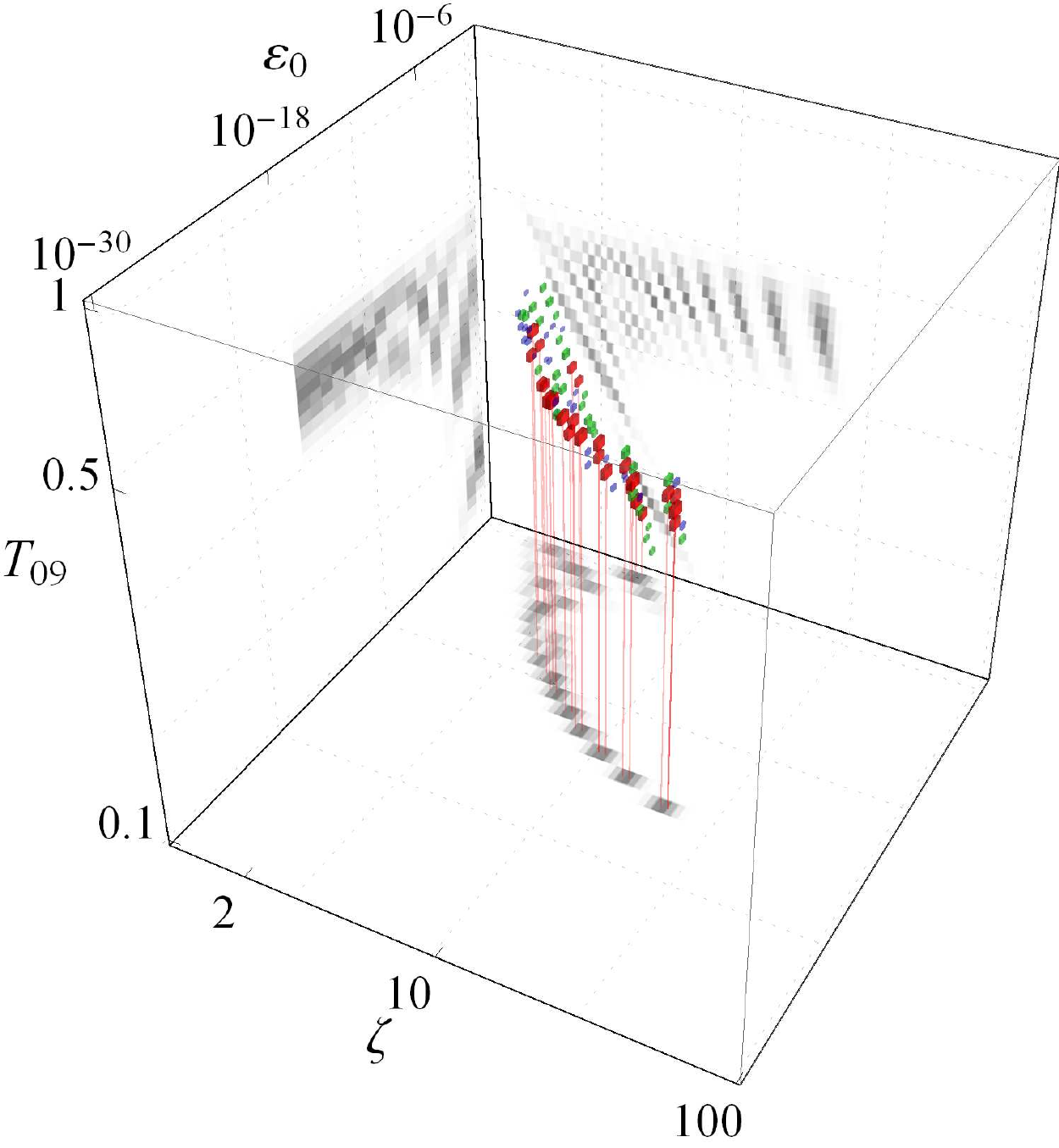} \caption{$\Delta=0.0625$} \end{subfigure}
\begin{subfigure}{.45\textwidth} \centering \includegraphics[width=0.95\textwidth]{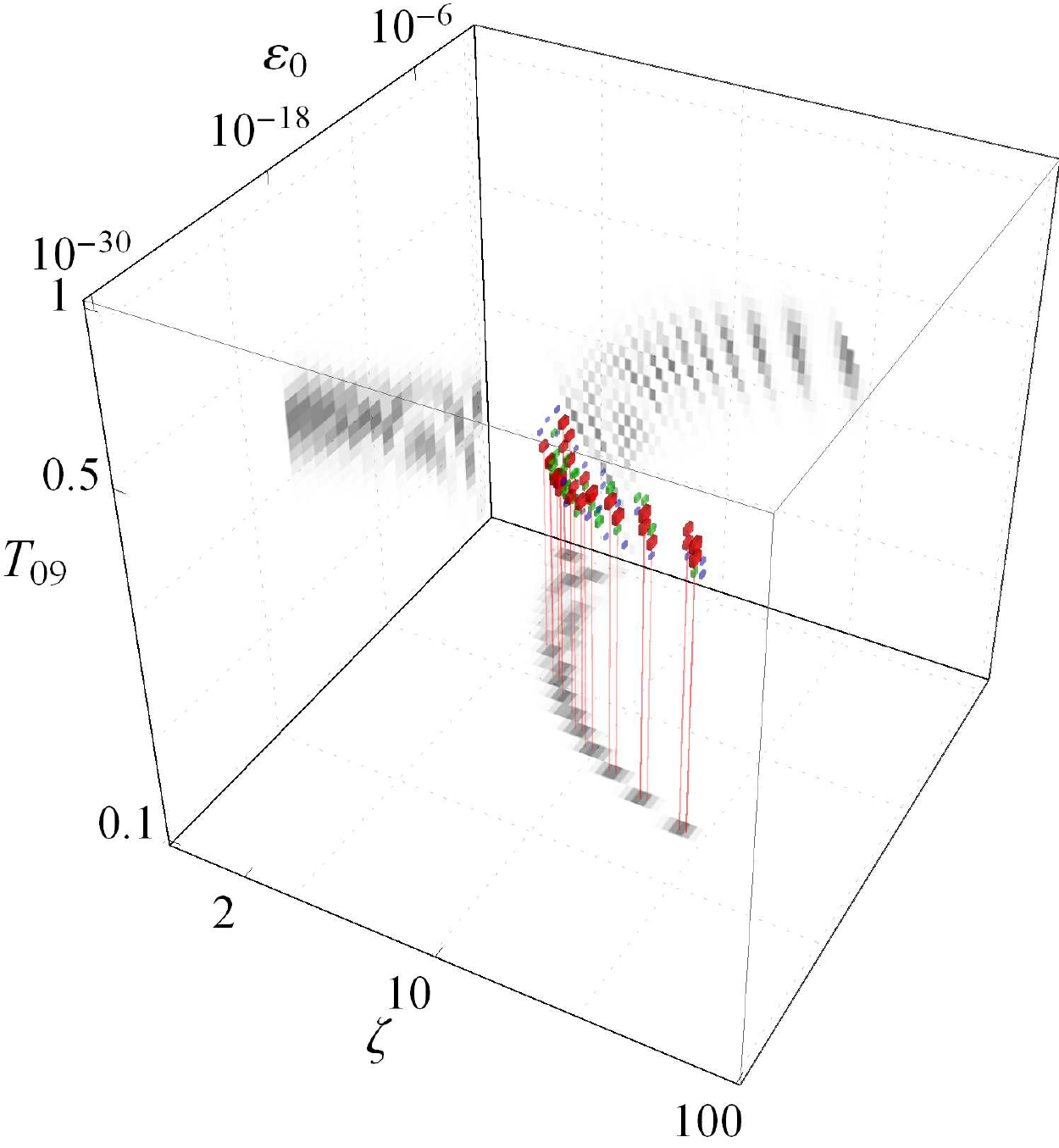} \caption{$\Delta=0.25$} \end{subfigure}
\\ \vskip 2mm 
\begin{subfigure}{.45\textwidth} \centering \includegraphics[width=0.95\textwidth]{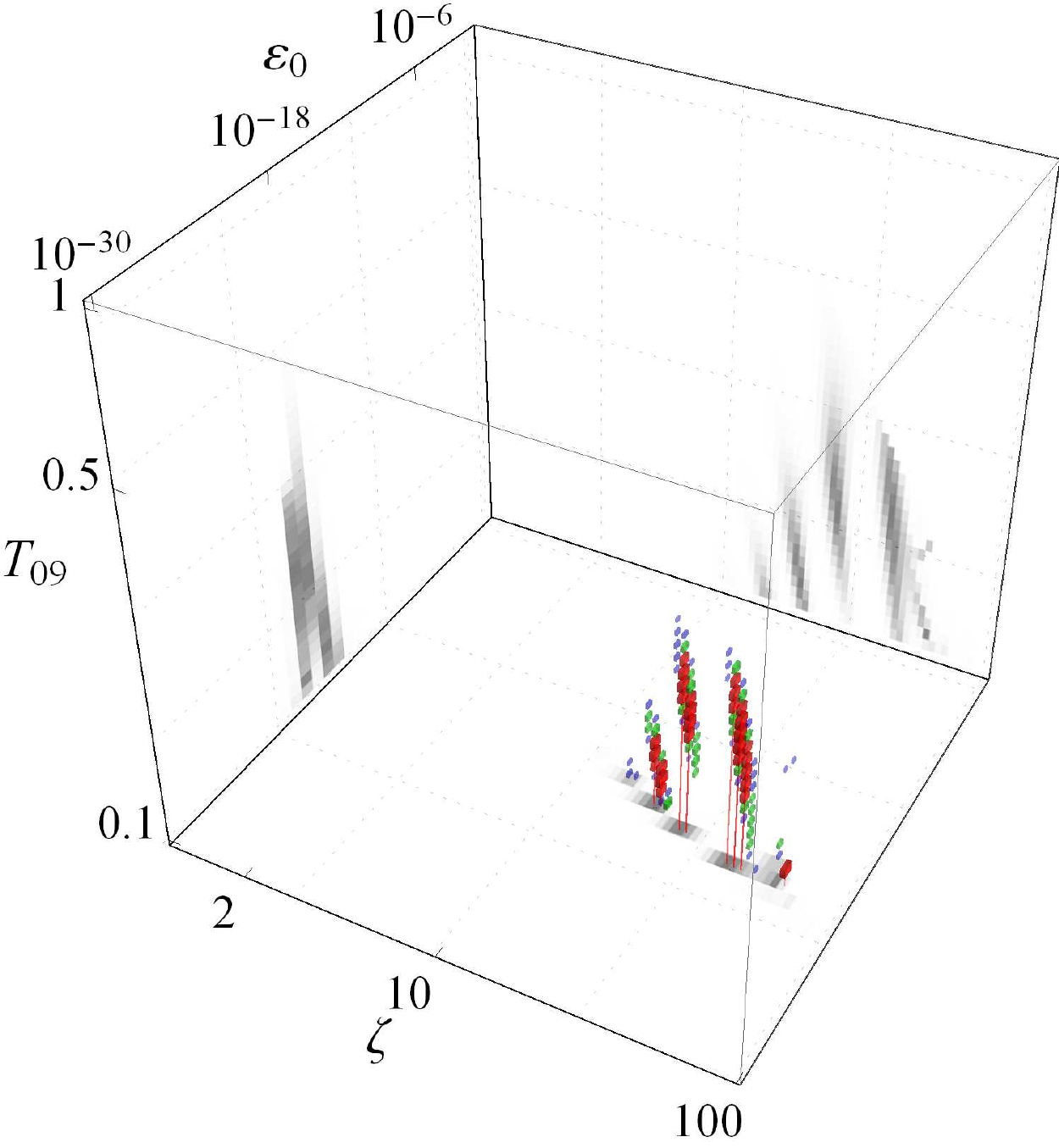} \caption{$\Delta=1$} \end{subfigure}
\begin{subfigure}{.45\textwidth} \centering \includegraphics[width=0.95\textwidth]{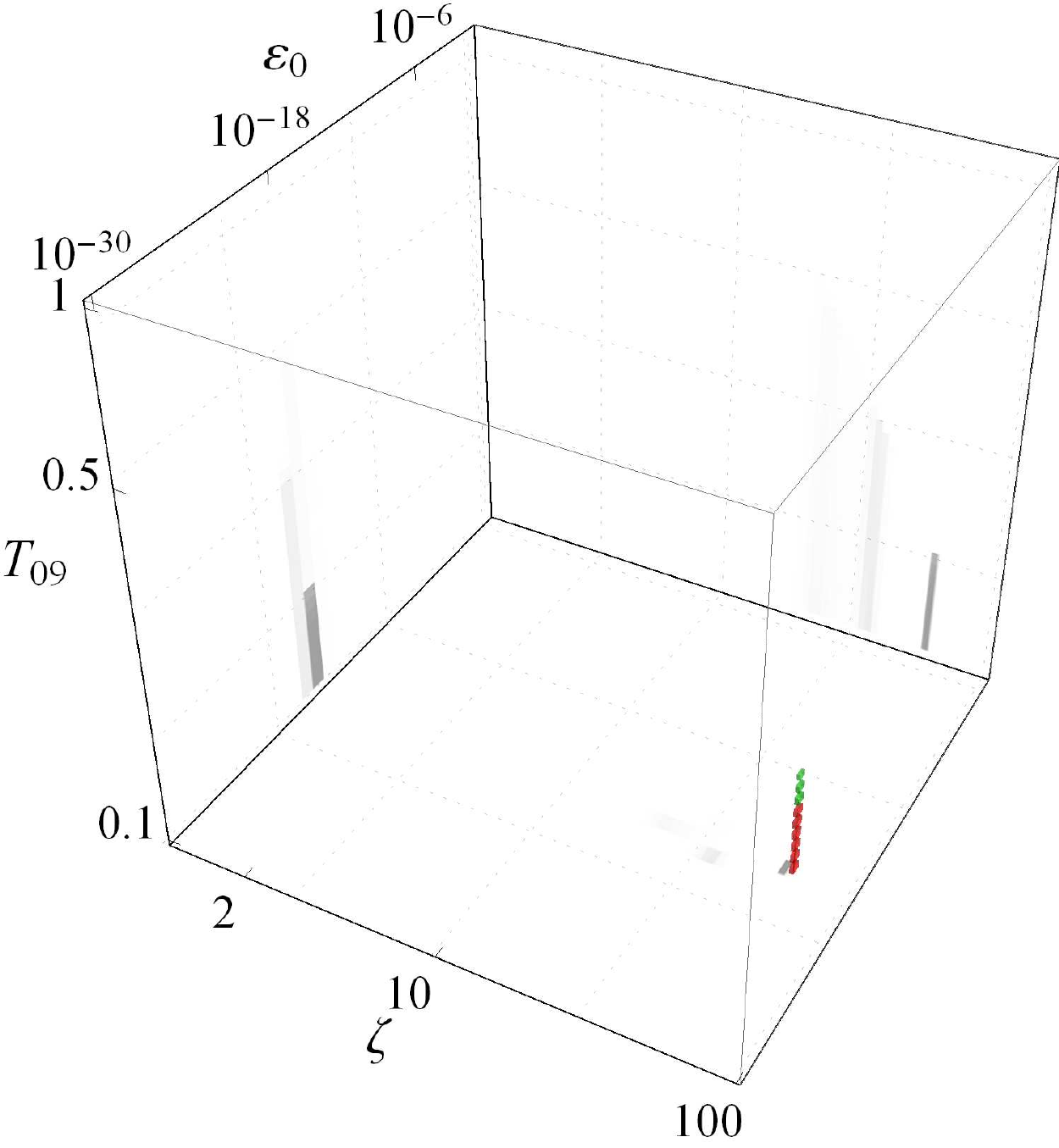} \caption{$\Delta=4$} \end{subfigure}
\caption{The $\chi^2$ values are plotted 
in the three-dimensional parameter space
$(\varepsilon_0, \zeta, T_{09})$ 
for $\Delta=0.0625$ (a), 0.25 (b), 1 (c) and 4 (d).
Though the grids are not shown here, 
the grid points
at which $\chi^2\le 3$ 
are represented by rectangular boxes,
whose sizes are proportional to the value of $e^{-\chi^2/3}$.
The boxes are colored according to the value of $\chi^2$:
Red for $\chi^2 \le 1$, 
green for $1 < \chi^2 \le 2$
and
blue for $2 < \chi^2 \le 3$.
For the cases where $\chi^2 \le 1$, thin lines are drawn 
to guide the eyes from the boxes down to the bottom plane.
The boxes are projected to the three planes 
($\varepsilon_0, \zeta), (\zeta, T_0)$, and ($T_0, \varepsilon_0$),
where a darker gray color corresponds to a smaller $\chi^2$.}
\label{wid-big}
\end{figure*}

The results of our calculations of $\chi^2$ 
are presented in the three-dimensional parameter space
$(\zeta,\, \varepsilon_0,\, T_{09})$
in Fig.~\ref{wid-big} for a few selected values of $\Delta$
by drawing rectangular boxes whose sizes are 
proportional to the value of 
$\exp(-\chi^2/3)$ at each grid site 
for the cases
when the value of 
$\chi^2 \le 3$.
The projections of the rectangular boxes are also
shown on the three planes in the parameter space $(\zeta,\, \varepsilon_0,\, T_{09})$.

Figure~\ref{wid-big}(a) shows that
when the width parameter is as small as $\Delta = 0.0625$,
the parameter set with a small $\chi^2$ 
are located in a narrow temperature range
with $0.37 \lesssim T_{09} \lesssim 0.43$.
This band structure
becomes irregular as $\Delta$ becomes bigger,
and then disappears for $\Delta \ge 1$.
We may understand this result as follows.
Equation~(\ref{window}) implies that
the magnitude of the non-thermal distribution $\varepsilon(T)$
is non-trivial
only during the period when the temperature
lies in
$T\simeq T_0 (e^{-\Delta} \sim e^{\Delta})$;
otherwise the magnitude is exponentially suppressed.
If $\Delta$ is small, the NTD is effective only in a narrow range
around $T\simeq T_0$,
and thus the $\chi^2$ minimum can be localized in the
$T_0$ space. 
The results with a small $\Delta$ in Fig.~\ref{wid-big}
imply that the NTD contribution in the narrow range 
around $T \simeq 0.40 \times 10^9 \mbox{K}$ 
plays the key role
in lowering $\chi^2$.
On the other hand, if $\Delta$ is large, 
$\varepsilon(T)$ becomes less sensitive to the $T_0$ parameter.  
As long as the ``window" of $\varepsilon(T)$
overlaps to some extent with the temperature range 
around $T \simeq 0.40 \times 10^9 \mbox{K}$, 
there is a potential to cure the lithium problem.
The estimation of the range of $T_{09}$ by using
$T_{09} \simeq 0.4 \times (e^{-\Delta} \sim e^{\Delta})$,
inferred from Eq.~(\ref{window}) gives us
$T_{09} \simeq (0.38 \sim 0.43)$ for $\Delta=0.0625$,
$T_{09} \simeq (0.31 \sim 0.51)$ for $\Delta=0.25$
and 
$T_{09} \simeq (0.15 \sim 1)$ for $\Delta=1$,
which are more or less consistent with Fig.~\ref{wid-big}.
Fig.~\ref{wid-big}(d) shows 
there are not many grid points 
where $\chi^2 \le 3$ when $\Delta = 4$.
As will be shown in subsection~\ref{small-chi2s},
we cannot find good parameter sets with $\chi^2 \le 3$ if $\Delta = 4$.
Thus, 
the width parameter $\Delta$ seems
to be limited 
to $\Delta = 1$ or less.

\begin{table*}
\centering
\caption{12 points in the parametric space which have $\chi^2 < 0.1$.
To get the values of $\chi^2$ per degree of freedom,
$1/3$ needs to be multiplied by the values 
in the Table.}\label{list-results-low}
\begin{tabular}{rlllllllllll}
\hline
No & $\chi^2$ & $\varepsilon_0$& $\Delta$ & $\zeta$ & $T_{09}$  & D/H         & T/H         & $\!^3$He/H  & $\!^6$Li/H   & $\!^7$Li/H   & $\!^7$Be/H\\
    &          &        &        &         & [$10^9$K] & [$10^{-5}$]  & [$10^{-8}$] & [$10^{-5}$] & [$10^{-14}$] & [$10^{-10}$] & [$10^{-10}$] \\
\hline
 1 & 0.006 & $10^{-12}\!$ & 0.2500 & 5.62  & 0.282 & 2.528 & 5.692 & 0.844 & 1.091 & 0.242 & 1.355\\
 2 & 0.011 & $10^{-15}\!$ & 0.5000 & 10.6  & 0.200 & 2.532 & 6.103 & 0.834 & 1.093 & 0.331 & 1.291\\
 3 & 0.015 & $10^{-19}\!$ & 1.0000 & 42.2  & 0.188 & 2.532 & 10.60 & 0.936 & 1.099 & 0.527 & 1.045\\
 4 & 0.032 & $10^{-16}\!$ & 0.1250 & 12.6  & 0.398 & 2.535 & 5.638 & 0.838 & 1.095 & 0.189 & 1.376\\
 5 & 0.046 & $10^{-12}\!$ & 0.0625 & 5.31  & 0.398 & 2.528 & 5.623 & 0.841 & 1.091 & 0.181 & 1.358\\    
 6 & 0.048 & $10^{-02}\!$ & 0.1250 & 2.82  & 0.224 & 2.535 & 6.482 & 0.956 & 1.095 & 0.458 & 1.093\\
 7 & 0.054 & $10^{-01}\!$ & 0.0625 & 4.47  & 0.141 & 2.527 & 14.40 & 1.021 & 1.195 & 0.402 & 1.262\\
 8 & 0.063 & $10^{-11}\!$ & 0.2500 & 5.01  & 0.266 & 2.522 & 5.683 & 0.842 & 1.089 & 0.246 & 1.315\\
 9 & 0.063 & $10^{-17}\!$ & 1.0000 & 20.0  & 0.112 & 2.521 & 7.534 & 0.823 & 1.089 & 0.434 & 1.186\\
10 & 0.064 & $10^{-19}\!$ & 1.0000 & 42.2  & 0.178 & 2.528 & 10.65 & 0.942 & 1.098 & 0.537 & 1.135\\
11 & 0.069 & $10^{-01}\!$ & 0.1250 & 2.11  & 0.282 & 2.539 & 5.821 & 0.950 & 1.097 & 0.188 & 1.379\\
12 & 0.077 & $10^{-16}\!$ & 0.5000 & 13.3  & 0.224 & 2.521 & 6.273 & 0.832 & 1.088 & 0.361 & 1.282\\
\hline
\end{tabular}
\end{table*}

\begin{table*}
\centering
\caption{The parameter sets with minimum $\chi^2$ for each value of $\varepsilon_0$.
   Only the cases with $\chi^2 < 10$ are listed.}\label{list-results123}
\begin{tabular}{lcccclllllll}
\hline\hline
$\varepsilon_0$& $\Delta$ & $\zeta$   & $T_{09}$ &   $\chi^2$ & D/H         & T/H         & $^3$He/H    & $^6$Li/H     & $^7$Li/H     & $^7$Be/H \\
            &        &      & [$10^9$K] &        &[$10^{-5}$]  & [$10^{-8}$] & [$10^{-5}$] & [$10^{-14}$] & [$10^{-10}$] & [$10^{-10}$] \\
\hline
$10^{-20}$  & 1 & 75.0  & 0.100 & 6.716 & 2.633 & 19.79 & 1.419  & 1.274 & 1.416 & 0.267\\
$10^{-19}$  & 1 & 42.2  & 0.188 & 0.015 & 2.532 & 10.60 & 0.936  & 1.099 & 0.527 & 1.045\\
$10^{-18}$  & $1/2$  & 25.1  & 0.316 & 0.100 & 2.532 & 6.373 & 0.815  & 1.094 & 0.350 & 1.158\\
$10^{-17}$  & 1 & 20.0  & 0.112 & 0.063 & 2.521 & 7.534 & 0.823  & 1.089 & 0.434 & 1.186\\
$10^{-16}$  & $1/8$ & 12.6  & 0.398 & 0.032 & 2.535 & 5.638 & 0.838  & 1.095 & 0.189 & 1.376\\
$10^{-15}$  & $1/2$ & 10.6  & 0.200 & 0.011 & 2.532 & 6.103 & 0.834  & 1.093 & 0.331 & 1.291\\
$10^{-14}$  & $1/16$ & 7.94  & 0.398 & 0.153 & 2.519 & 5.596 & 0.833  & 1.088 & 0.182 & 1.336\\
$10^{-13}$  & $1/16$ & 6.31  & 0.398 & 0.235 & 2.539 & 5.669 & 0.861  & 1.096 & 0.189 & 1.537\\
$10^{-12}$  & $1/4$ & 5.62  & 0.282 & 0.006 & 2.528 & 5.692 & 0.844  & 1.091 & 0.242 & 1.355\\
$10^{-11}$  & $1/4$ & 5.01  & 0.266 & 0.063 & 2.522 & 5.683 & 0.842  & 1.089 & 0.246 & 1.315\\
$10^{-10}$  & $1/2$ & 5.96  & 0.100 & 0.142 & 2.519 & 5.995 & 0.845  & 1.088 & 0.335 & 1.341\\
$10^{-09}$  & $1/8$ & 3.55  & 0.355 & 0.148 & 2.536 & 5.657 & 0.852  & 1.095 & 0.190 & 1.305\\
$10^{-08}$  & $1/8$ & 3.16  & 0.355 & 0.317 & 2.546 & 5.685 & 0.859  & 1.100 & 0.186 & 1.299\\
$10^{-07}$  & $1/8$ & 2.99  & 0.335 & 0.151 & 2.545 & 5.711 & 0.880  & 1.099 & 0.200 & 1.376\\
$10^{-06}$  & $1/16$ & 2.66  & 0.376 & 0.330 & 2.549 & 5.724 & 0.890  & 1.101 & 0.189 & 1.317\\
$10^{-05}$  & $1/16$ & 2.99  & 0.299 & 1.351 & 2.538 & 5.947 & 0.939  & 1.096 & 0.412 & 1.532\\   
$10^{-04}$  & $1/8$  & 2.66  & 0.282 & 0.201 & 2.540 & 5.804 & 0.918  & 1.097 & 0.271 & 1.218\\
$10^{-03}$  & $1/4$  & 2.82  & 0.178 & 0.345 & 2.535 & 5.774 & 0.887  & 1.096 & 0.248 & 1.181\\
$10^{-02}$  & $1/8$  & 2.82  & 0.224 & 0.048 & 2.535 & 6.482 & 0.956  & 1.095 & 0.458 & 1.093\\
$10^{-01}$  & $1/16$  & 4.47  & 0.141 & 0.054 & 2.527 & 14.40 & 1.021  & 1.195 & 0.402 & 1.262\\
\hline
\end{tabular}
\end{table*}

\begin{figure}
\centering
\includegraphics[width=0.6\textwidth]{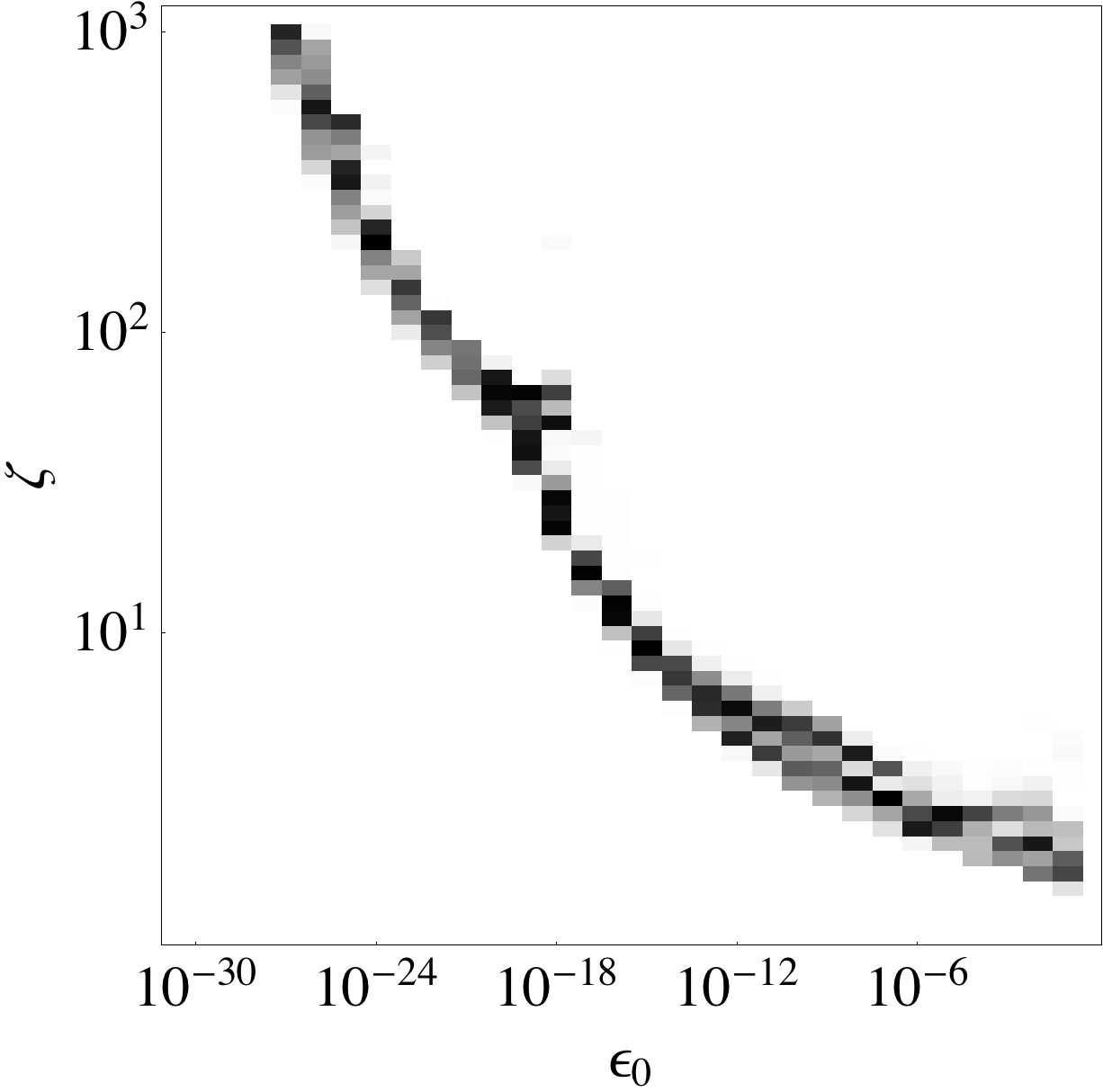}
\caption{
Projection of $\exp(-\chi^2/3)$ on the $(\varepsilon_0,\,\zeta)$ 
plane,
while the value of $\Delta$ is adjusted to yield the minimum
$\chi^2$ at each grid site.
}
\label{zetaeps0-best}
\end{figure}

As can be seen from the projections plotted
on the bottom plane in Fig.~\ref{wid-big},
the parameter $\zeta = T_{\rm NTD}/T$ turns out to be 
strongly correlated with $\varepsilon_0$.
This is natural because
the amount of the NTD portion required to cure the lithium problem would be smaller
if the temperature of the NTD becomes higher.
It is noteworthy that the correlation curves 
are quite insensitive to the value of $\Delta$,
which can be seen by comparing the projections in the ($\varepsilon_0 , \zeta$) plane 
for different values of $\Delta$ in 
Fig.~\ref{wid-big}.
The projection of $\exp(-\chi^2/3)$ on the $(\varepsilon_0,\, \zeta)$ plane
is displayed in 
Fig.~\ref{zetaeps0-best},
where 
the value of $\Delta$ is adjusted to yield the minimum
$\chi^2$ at each grid site.
If $\zeta$ is chosen as 20, 10 and 5,
the required fraction to yield a small $\chi^2$ becomes
$\varepsilon_0 \simeq 10^{-17}$, $10^{-15}$ and $10^{-11}$, respectively.

\begin{figure*}
\centering
\begin{subfigure}{.45\textwidth} \centering \includegraphics[width=0.95\textwidth]{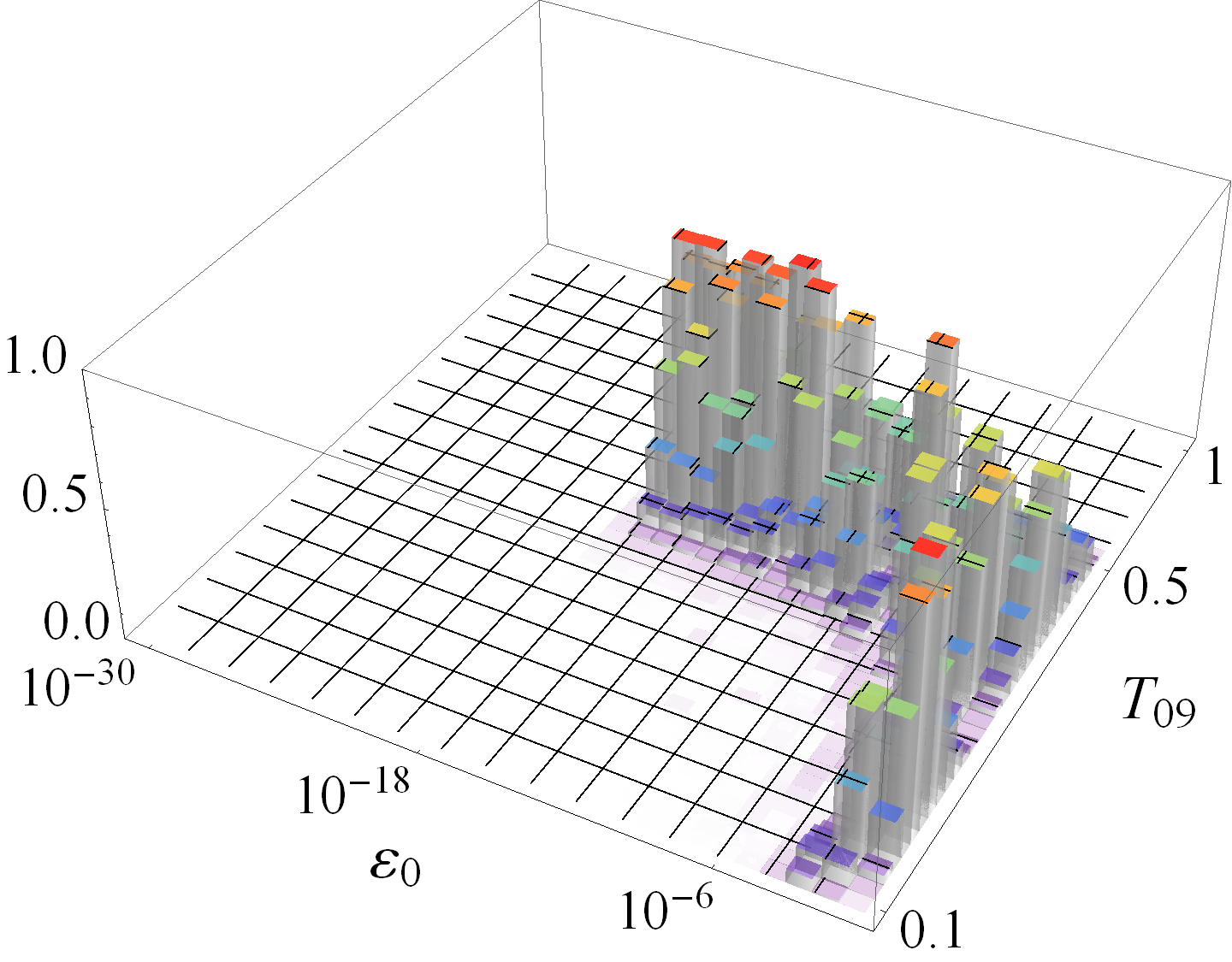} \caption{$\Delta=0.0625$} \end{subfigure}
\begin{subfigure}{.45\textwidth} \centering \includegraphics[width=0.95\textwidth]{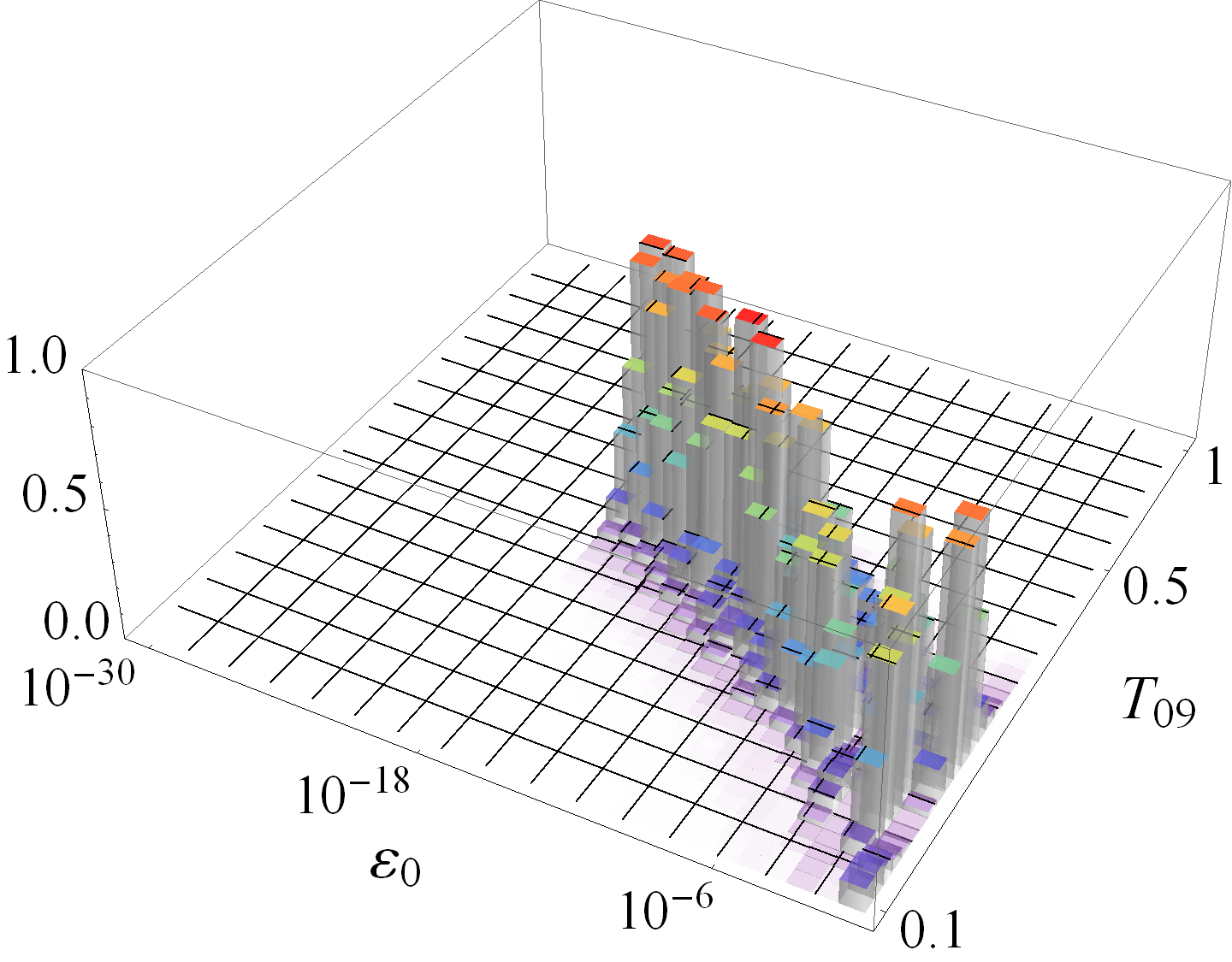} \caption{$\Delta=0.25$} \end{subfigure}
\\ \vskip 5mm 
\begin{subfigure}{.45\textwidth} \centering \includegraphics[width=0.95\textwidth]{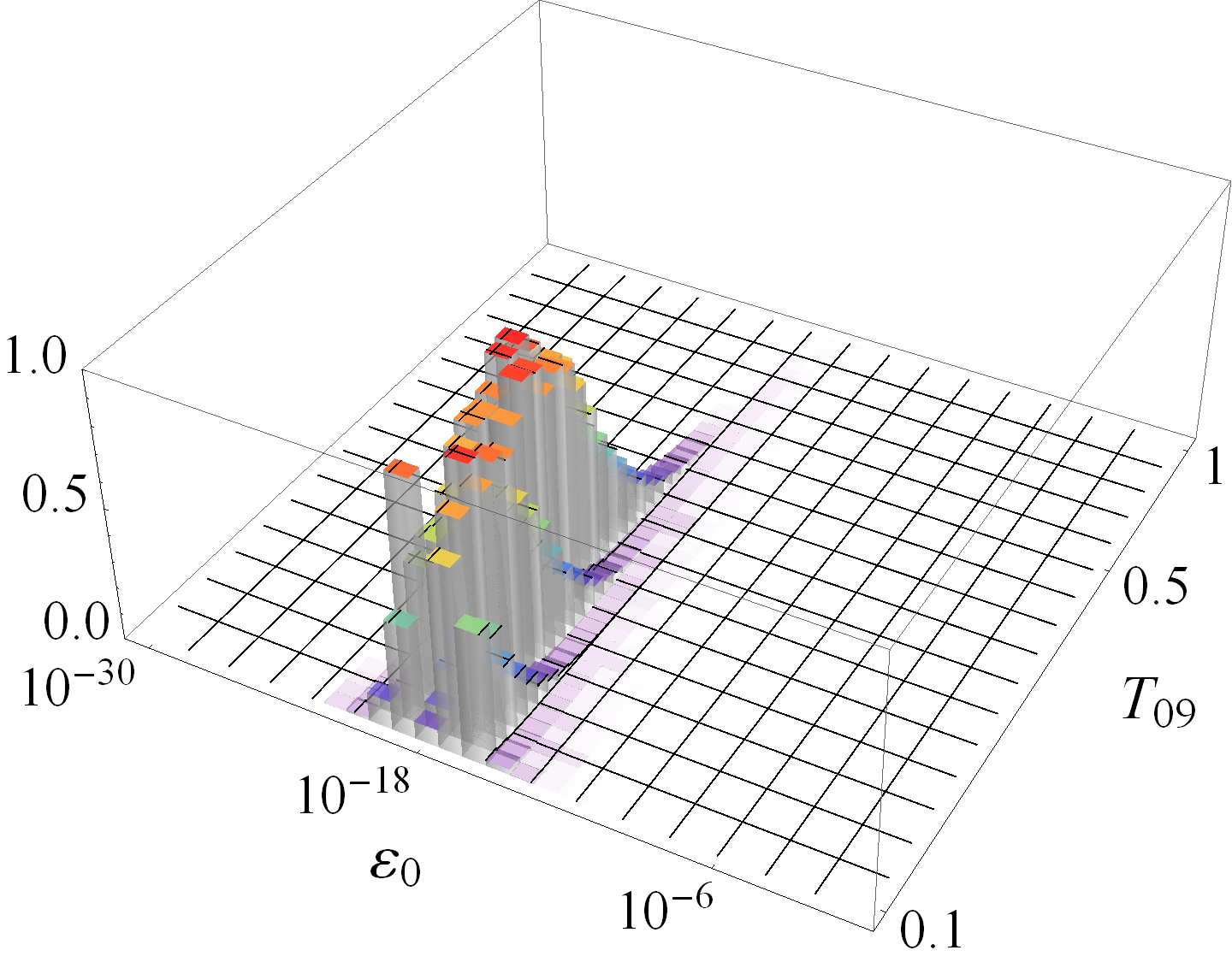} \caption{$\Delta=1$} \end{subfigure}
\begin{subfigure}{.45\textwidth} \centering \includegraphics[width=0.95\textwidth]{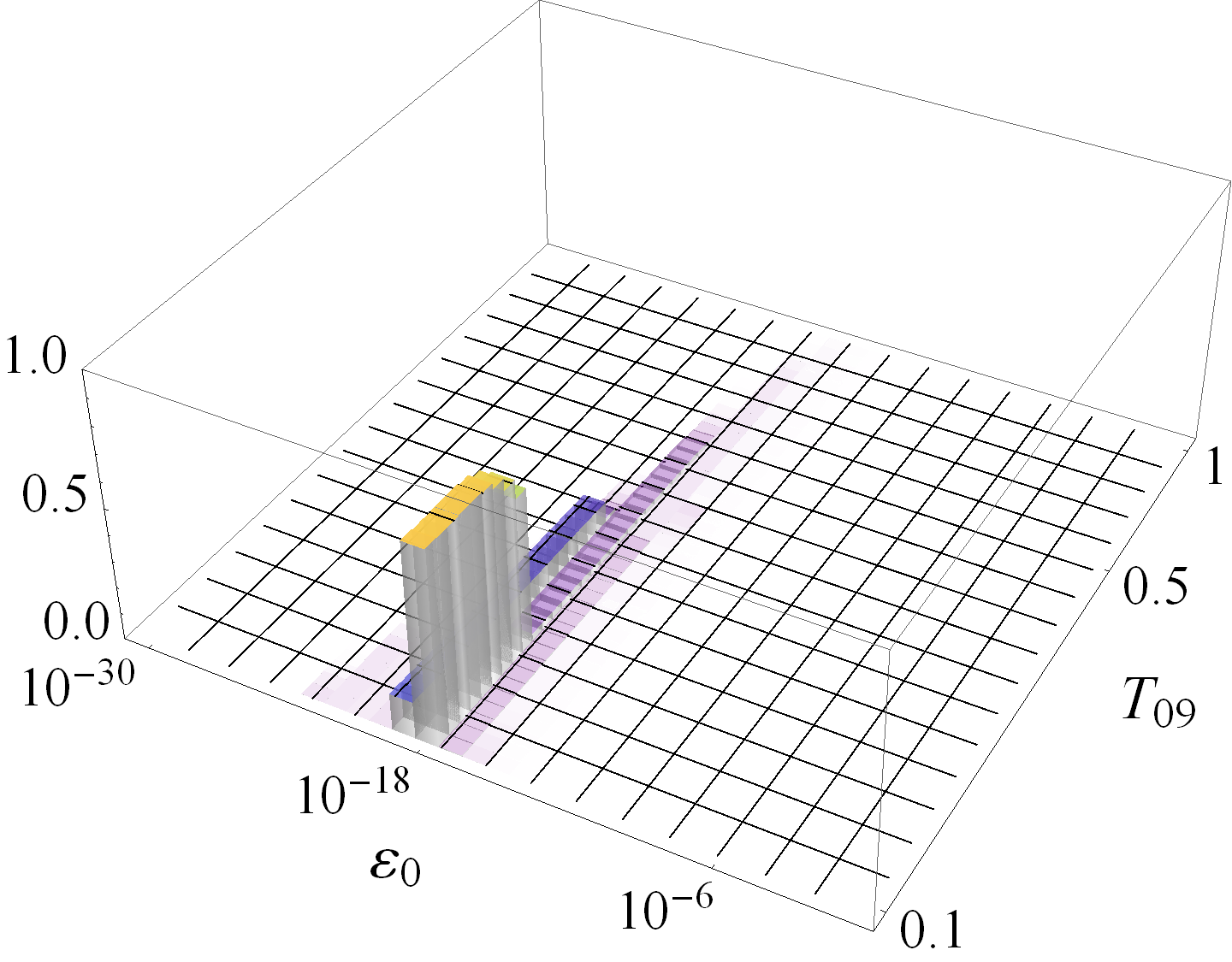} \caption{$\Delta=4$} \end{subfigure}
\caption{The value of $\exp(-\chi^2/3)$ on the $(\varepsilon_0,\, T_{09})$ plane,
where the width parameter is $\Delta=0.0625$ (a), $0.25$ (b), 1 (c) and 4 (d).
The value of $\zeta$ is adjusted 
to yield the minimum $\chi^2$ at each grid point.}
\label{epsT0II}
\end{figure*}

In Fig.~\ref{epsT0II},
$\exp(-\chi^2/3)$ is plotted in the $(\varepsilon_0,\,T_{09})$ plane
for a few selected values of $\Delta$.
At each grid point,
the value of $\zeta$ is further adjusted to produce the minimum $\chi^2$
while its rough value can be inferred
from the aforementioned correlation curve 
between $\varepsilon_0$ and $\zeta$.
Figure \ref{epsT0II} demonstrates, in particular, 
how the band-type structure
observed
around $T_{09} \simeq 0.40$
for a small $\Delta$ evolves 
as $\Delta$ becomes large.
For a width parameter $\Delta$ in the medium range, such as $\Delta=0.25$,
the band is formed in 
the diagonal direction in the ($\varepsilon_0$, $T_{09}$) plane,
while the band is rotated to
the direction of constant $\varepsilon_0$ $\sim 10^{-18}$
for $\Delta=1$ or larger.

\subsection{Parameter sets with small values of $\chi^2$\label{small-chi2s}}

More than 130 grid points are found to yield $\chi^2$ less than 1.
Among them, 12 cases have $\chi^2 < 0.1$,
which are listed
in Table~\ref{list-results-low}
with the resulting abundances of light elements.
The number of minima itself is not meaningful since it
will become larger if we take a finer mesh.
In Table~\ref{list-results123},
the values of $\Delta, \zeta$ and $T_{09}$
which minimize $\chi^2$ 
for each value of $\varepsilon_0$ considered here
are presented together with the abundances of elements.
There is no parameter set resulting in $\chi^2 < 10$ for
$\varepsilon_0 \le 10^{-21}$,
and 
for $\varepsilon_0 \ge 10^{-19}$
the parameter sets with $\chi^2<1$
are found.

In all the cases listed,
the proton and the $^4$He abundances
are found to be 
0.753 and 0.247, respectively.
The abundance of $^6$Li/H is quite insensitive to the value of $\varepsilon_0$,
resulting in $^6\mbox{Li/H} \simeq 1.1 \times 10^{-14}$.
The sum of 
abundances of
$^7$Li and $^7$Be is constrained by the
$\chi^2$ calculation,
and is 
consistent with the measured primordial 
lithium abundance
for all the parameter sets with small $\chi^2$ values listed in Table 2,
while there are
sizable variations in 
the abundances of
$^7$Li and $^7$Be.

\subsection{Evolution of abundances}

\begin{figure*}
\centering
%
\begin{subfigure}{\textwidth} \centering \includegraphics[width=0.45\textwidth]{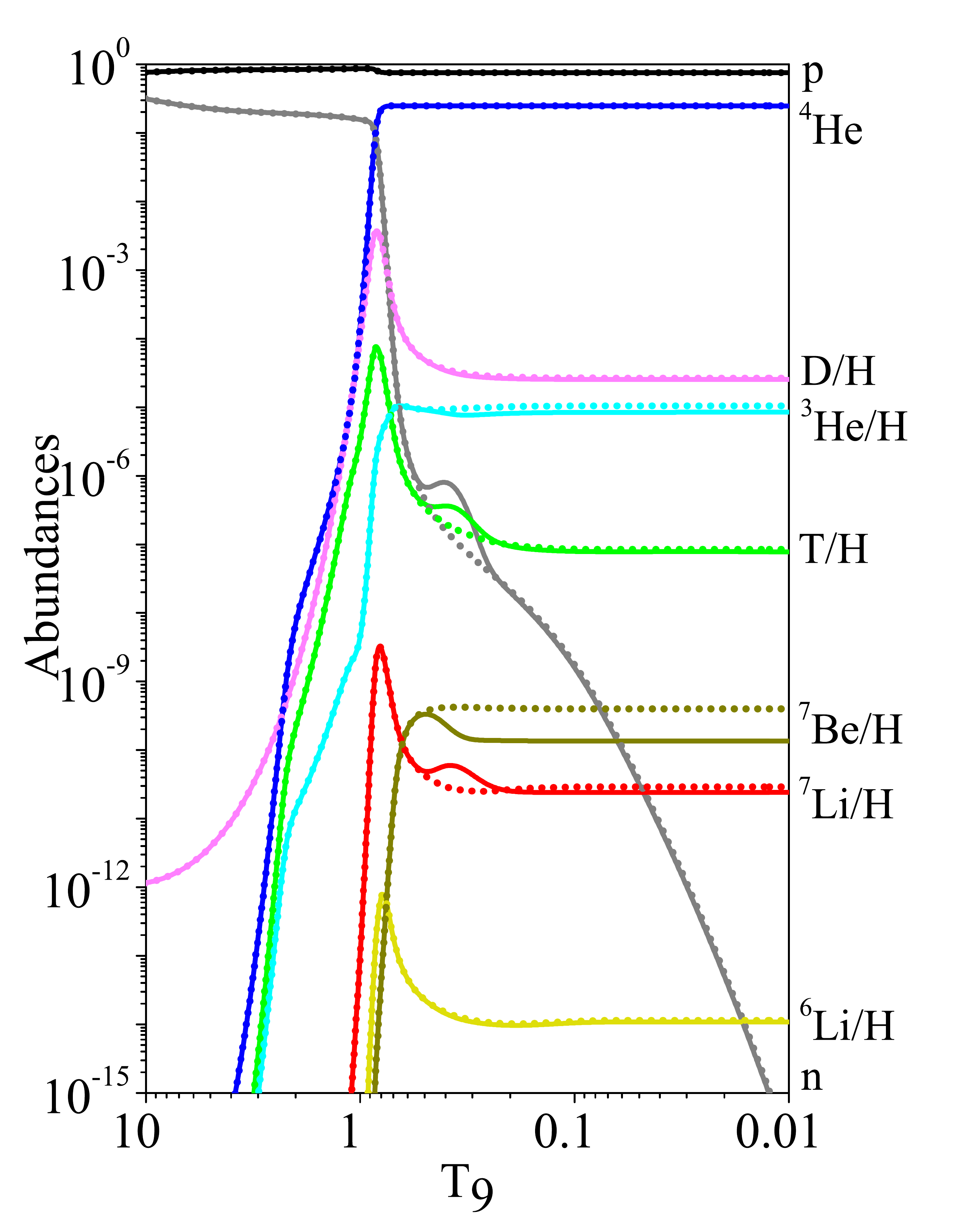} 
                                         \includegraphics[width=0.45\textwidth]{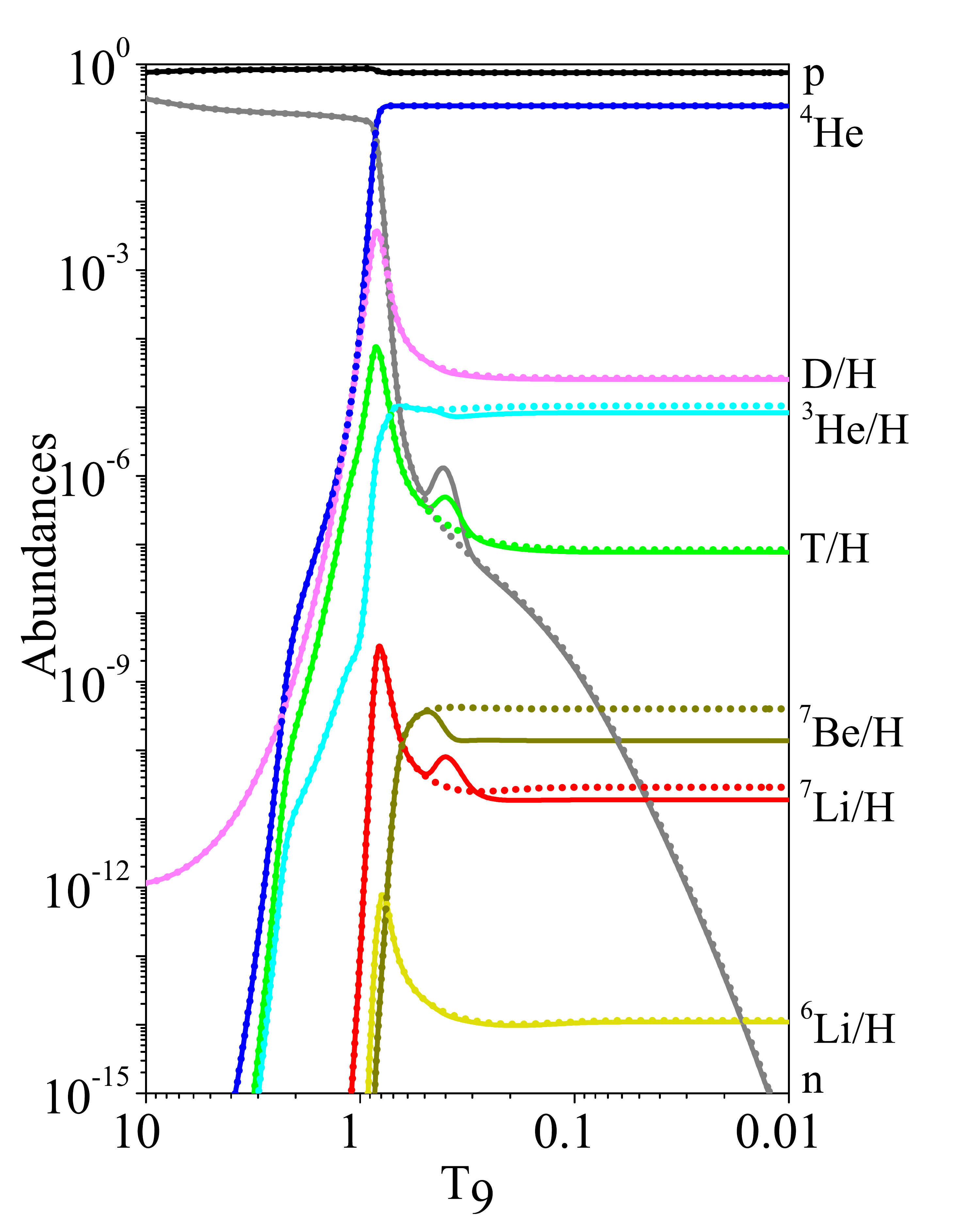} \end{subfigure}
\\ \vskip 5mm 
\begin{subfigure}{\textwidth} \centering \includegraphics[width=0.45\textwidth]{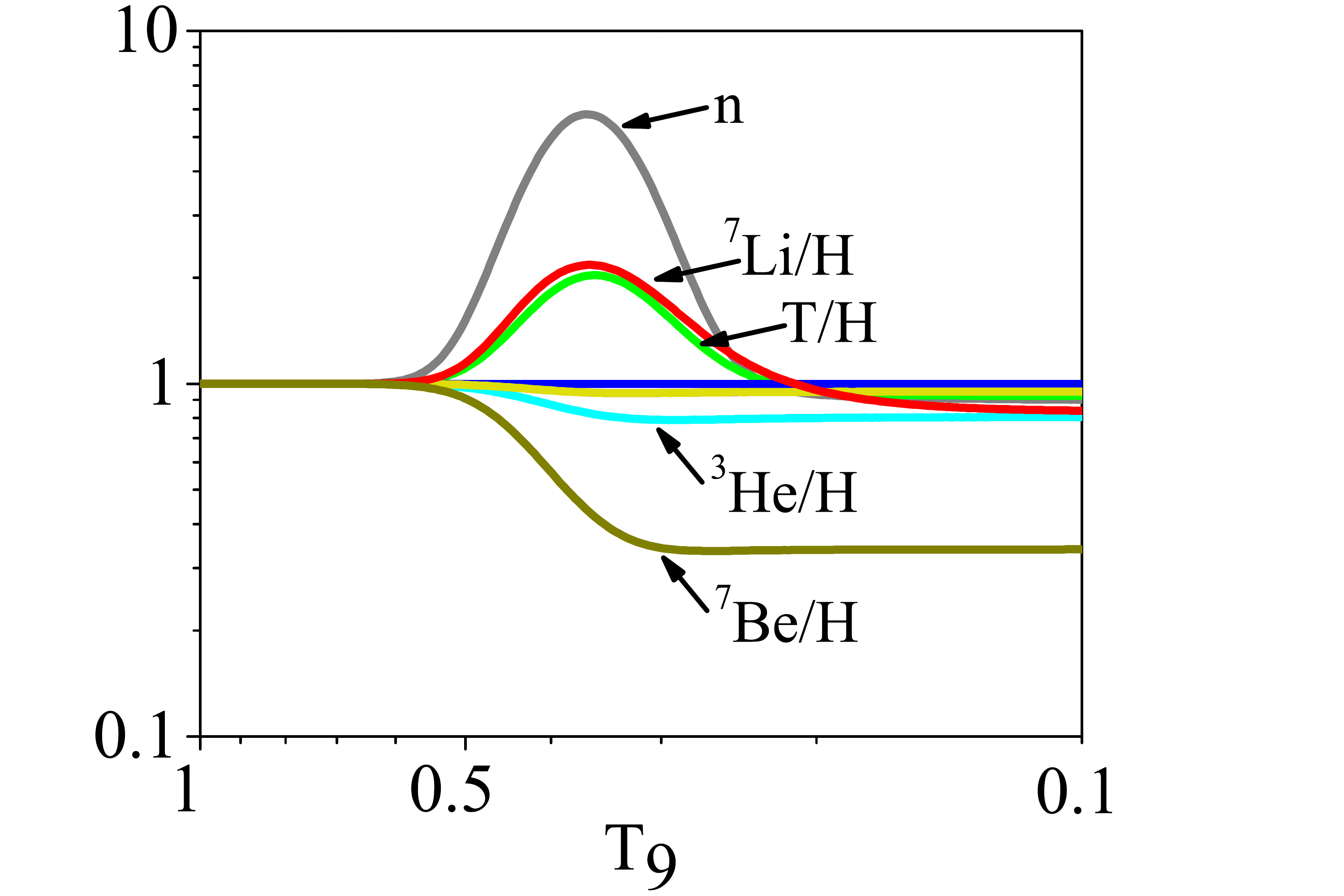} 
                                         \includegraphics[width=0.45\textwidth]{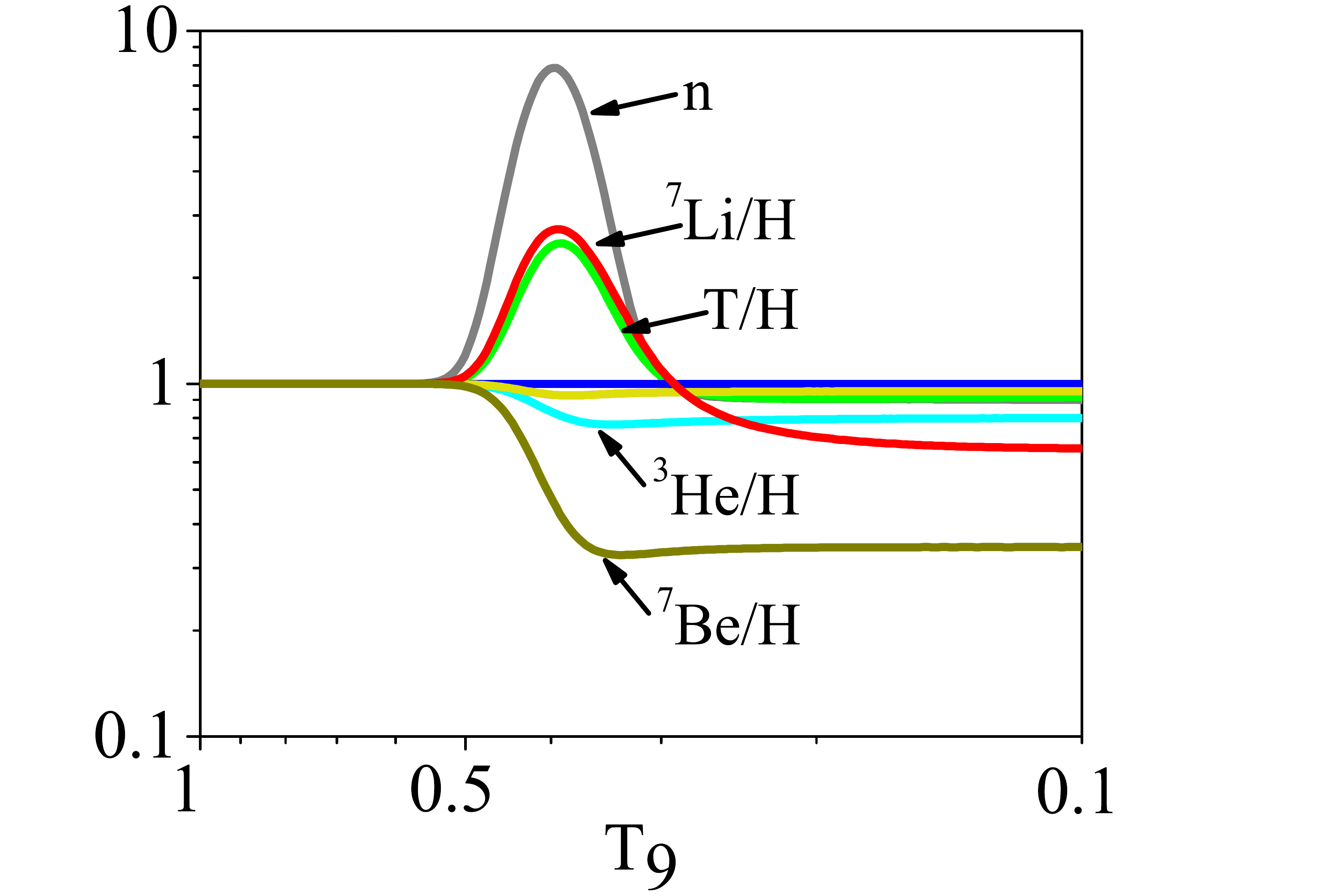} \end{subfigure}
\caption{
The abundances of elements
with (the solid lines) and without (the dotted lines) the NTD effects
are shown in the top panels,
and the relative ratios of the 
abundances of elements with the NTD effects
with respect to those without
are plotted in the bottom panels.
The left panels are for 
the parameter set No. 1 ($\varepsilon_0 = 10^{-12}$, $\Delta = 0.25$, $\zeta = 5.62$ and $T_0 = 0.282$),
and the right panels are for 
the set No. 4 ($\varepsilon_0 = 10^{-16}$, $\Delta = 0.125$, $\zeta = 12.6$ and $T_0 = 0.398$)
of Table~\ref{list-results-low}.}
\label{abun}
\end{figure*}

The evolution of abundance of light elements 
is shown for two selected cases in Fig.~\ref{abun}
with (the solid lines) and without (the dotted lines) 
taking into account the NTD contributions. 
The left panels are for the set No.~1 in Table~\ref{list-results-low} 
with $\chi^2 = 0.006$,
and the right panels are for the set No.~4 
with $\chi^2= 0.032$.
In the bottom panels,
the relative ratios of the abundances
with and without the NTD particles are plotted.

Several remarks are in order here.
First, major changes occur during the period
$0.5 \lesssim T_9 \lesssim 0.3$.
Secondly,
the NTD contribution is found to
lower the abundances of \atom{He}{}{3} and \atom{Be}{}{7}.
The neutron, triton and \atom{Li}{}{7} abundances
increase 
during this period
until $T_9 \simeq 0.35$ and then slightly decrease.
\atom{Be}{}{7} is dominant over \atom{Li}{}{7}, and
the sum of the two is lowered by about one-third, 
as required by the observation.
Thirdly,
as the neutron abundance increases around $T_9 = (0.5 \sim 0.3)$
due to the NTD contribution,
Coc et al.\cite{coc} 
have also concluded that
the injection of extra neutrons at $T\simeq 50$ keV (or $T_9 \simeq 0.43$)
can resolve the lithium problem
but at the cost of over-predicting the deuteron abundance
far beyond the observed values.

\begin{figure*}
\centering
%
\begin{subfigure}{\textwidth} \centering \includegraphics[width=0.45\textwidth]{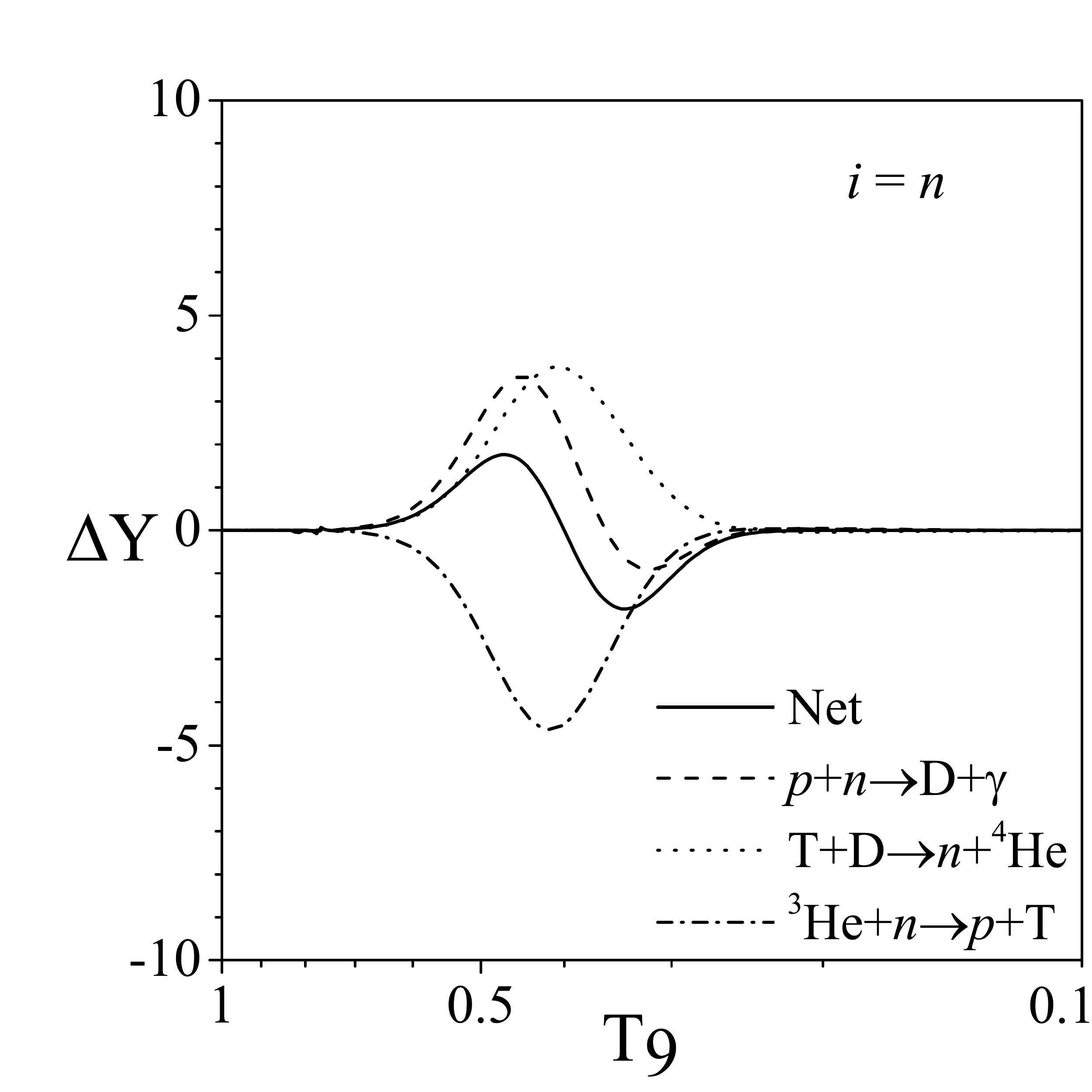} 
\includegraphics[width=0.45\textwidth]{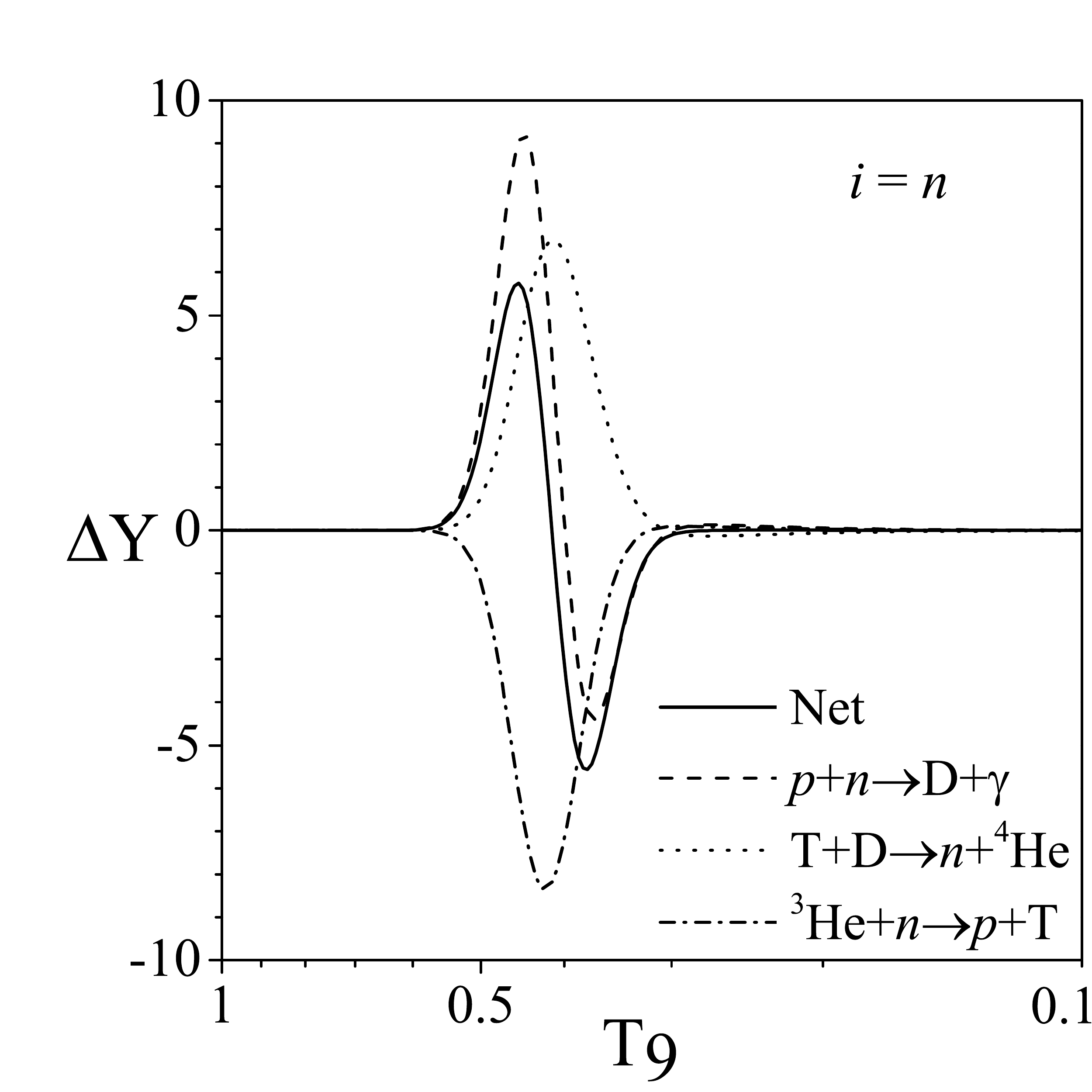}  \caption{neutron} \end{subfigure}
\\ 
\vskip 5mm
\begin{subfigure}{\textwidth} \centering \includegraphics[width=0.45\textwidth]{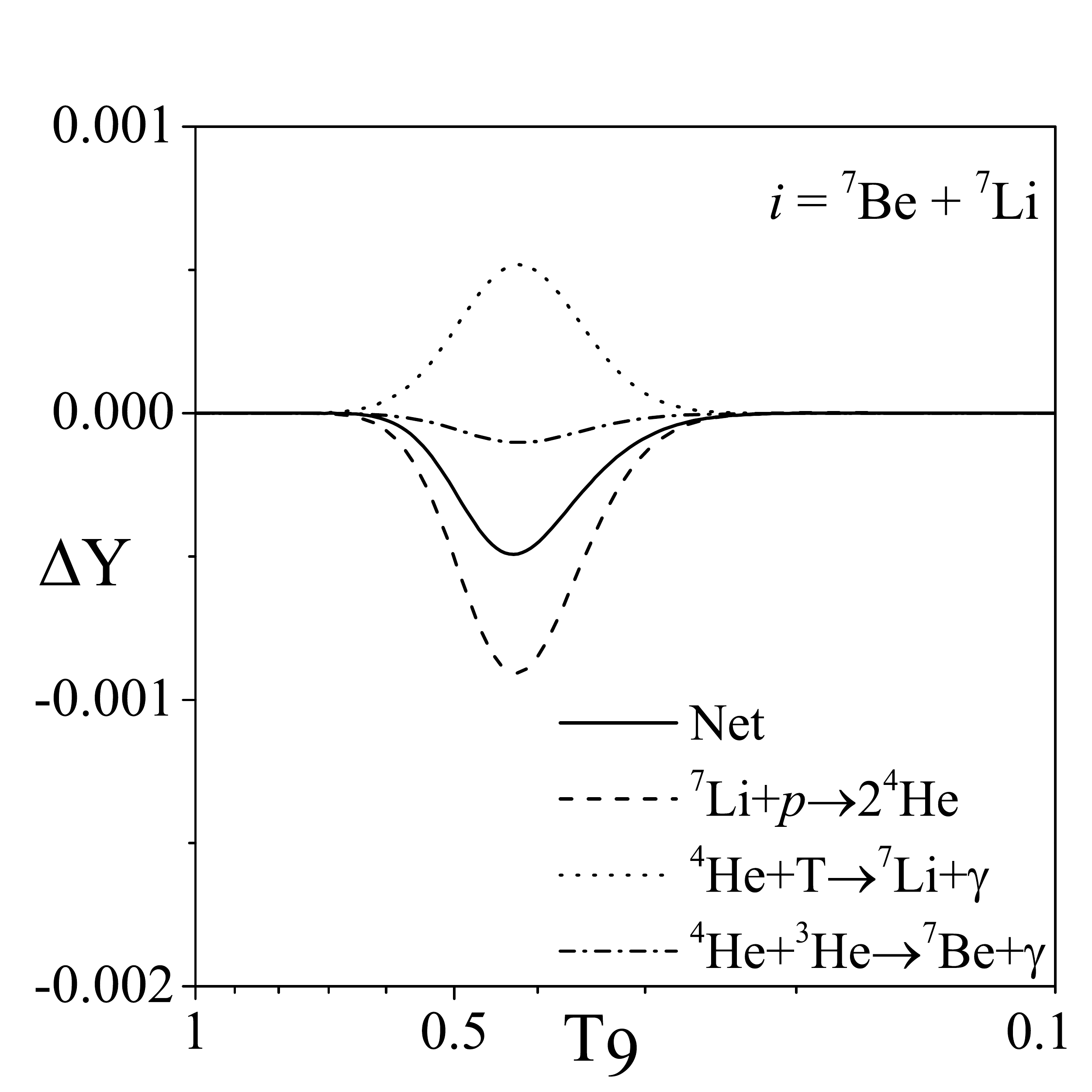} 
\includegraphics[width=0.45\textwidth]{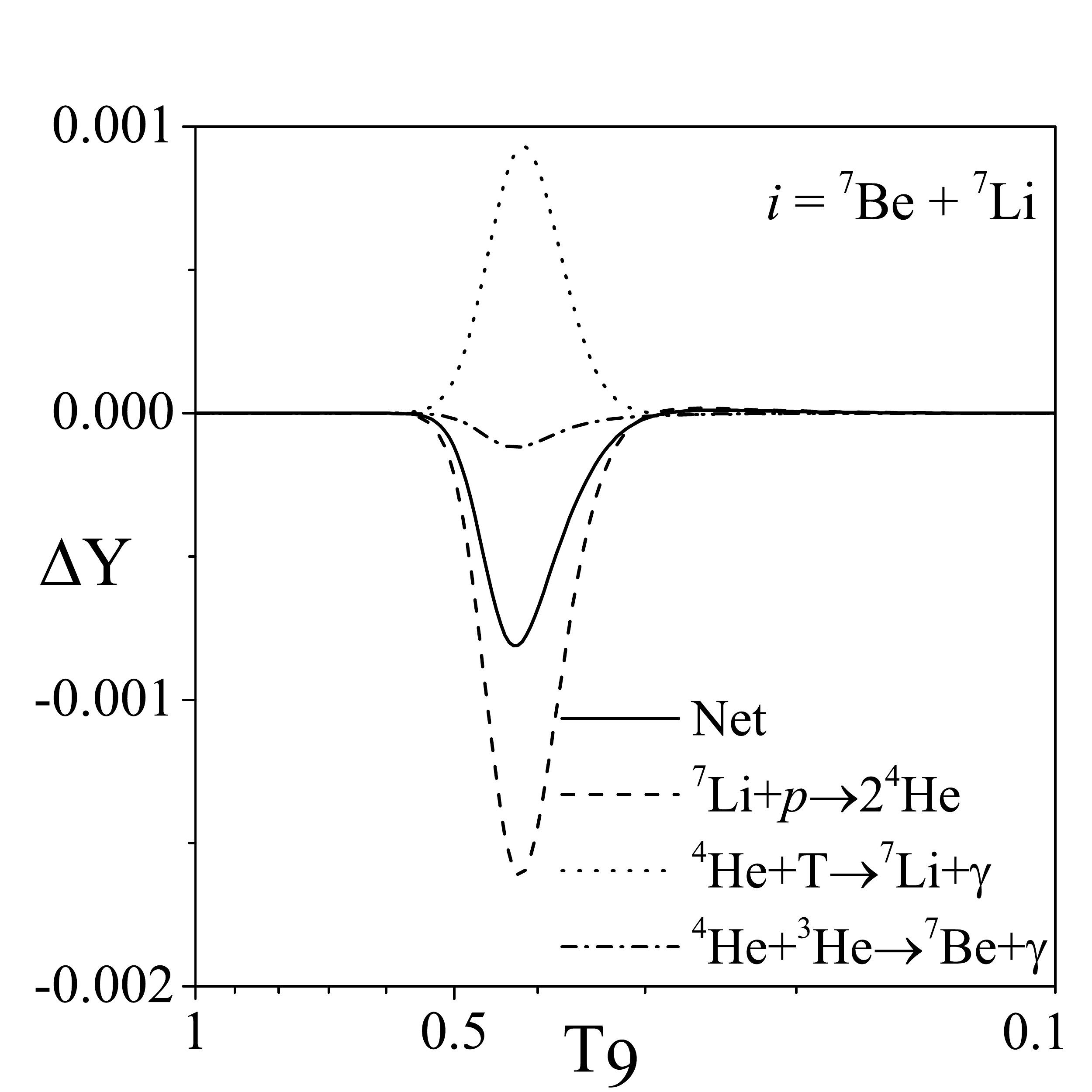}  \caption{$\atom{Be}{}{7}+\atom{Li}{}{7}$} \end{subfigure}
\caption{
The NTD induced changes in the production rate, Eq.~(\ref{diff-def}),
for neutrons (top) and $\atom{Be}{}{7}+\atom{Li}{}{7}$ (bottom).
The left and right panels are for
the the parameter set No. 1 and No. 4 of Table~\ref{list-results-low}, respectively.}
\label{reactionrate}
\end{figure*}

In Fig.~\ref{reactionrate},
we show the changes in the production rate
of light
elements due to the NTD contribution 
\begin{equation}
\Delta Y_i \equiv
\frac{1}{H}\frac{d}{d t} \left[ Y_i^{ntd} - Y_i^{th} \right],
\label{diff-def}
\end{equation}
where $Y_i^{ntd}$ and $Y_i^{th}$ denote the abundance of
the neutron ($i=n$) as shown in Fig.~\ref{reactionrate}($a$)
and that of the sum of $\atom{Be}{}{7}$ and $\atom{Li}{}{7}$ 
($i=\atom{Be}{}{7}+\atom{Li}{}{7})$ as shown in Fig.~\ref{reactionrate}($b$)
with and without the NTD contribution, respectively,
and $H$ is the Hubble parameter.
The left and right panels are
for the parameter sets No. 1 and No. 4 of Table~\ref{list-results-low}, respectively.
In addition to 
the net change (denoted by the solid lines),
we have also plotted the contributions from a few important reactions,
i. e., 
$p+n\to \atom{D}{}{}+\gamma$, 
$\atom{T}{}{}+\atom{D}{}{}\to n+\atom{He}{}{4}$, 
$\atom{He}{}{3}+n\to p+\atom{T}{}{}$,
$\atom{Li}{}{7}+p\to 2 \atom{He}{}{4}$
and
$\atom{He}{}{4}+\atom{T}{}{}\to \atom{Li}{}{7}+\gamma$.
The general behaviors of the curves from both sets
No.~1 and No.~4 are quite similar.
We observe that
the NTD component enhances the photo-disintegration of the deuteron
(the backward reaction  of $p+n\to \atom{D}{}{}+\gamma$) 
and $\atom{T}{}{}+\atom{D}{}{}\to n+\atom{He}{}{4}$ reaction,
increasing the neutron abundance.
On the other hand,
$\atom{He}{}{3}+n\to p+\atom{T}{}{}$ reaction is also
enhanced by the NTD, 
and 
reduces
the neutron abundance.
The bottom panels of Fig.~\ref{abun} also show
the neutron abundance
increases
until $T_{09} \simeq 0.4$
as much as about 8 times of the SBBN value,
and then
decreases to the level of
0.9 times of the SBBN value.

The change in $\atom{Li}{}{7}+\atom{Be}{}{7}$ abundance
is found to be dominated by
the NTD contribution 
from
the $\atom{Li}{}{7}+p\to 2 \atom{He}{}{4}$ reaction.
A high NTD temperature
enhances the burning of \atom{Li}{}{7}
by allowing the proton to overcome the 
Coulomb barrier and to fuse with
\atom{Li}{}{7}.
Roughly a half of the
reduction of Li 
due to this process is offset
by the increase of 
\atom{Li}{}{7} abundance through the
$\atom{He}{}{4}+\atom{T}{}{}\to \atom{Li}{}{7}+\gamma$ reaction.
Due to the increased neutron abundance,
the $\atom{Be}{}{7} + n \to p + \atom{Li}{}{7}$
reaction also increases the lithium abundance
roughly as much as 
$\atom{He}{}{4}+\atom{T}{}{}\to \atom{Li}{}{7}+\gamma$ reaction,
but it lowers the same amount of \atom{Be}{}{7}
leaving the sum of $\atom{Li}{}{7}+\atom{Be}{}{7}$ unchanged.
After summing the contributions from all the reactions,
the lithium abundance is lowered
to the measured value.

\begin{figure}
\centering
\includegraphics[width=0.8\textwidth]{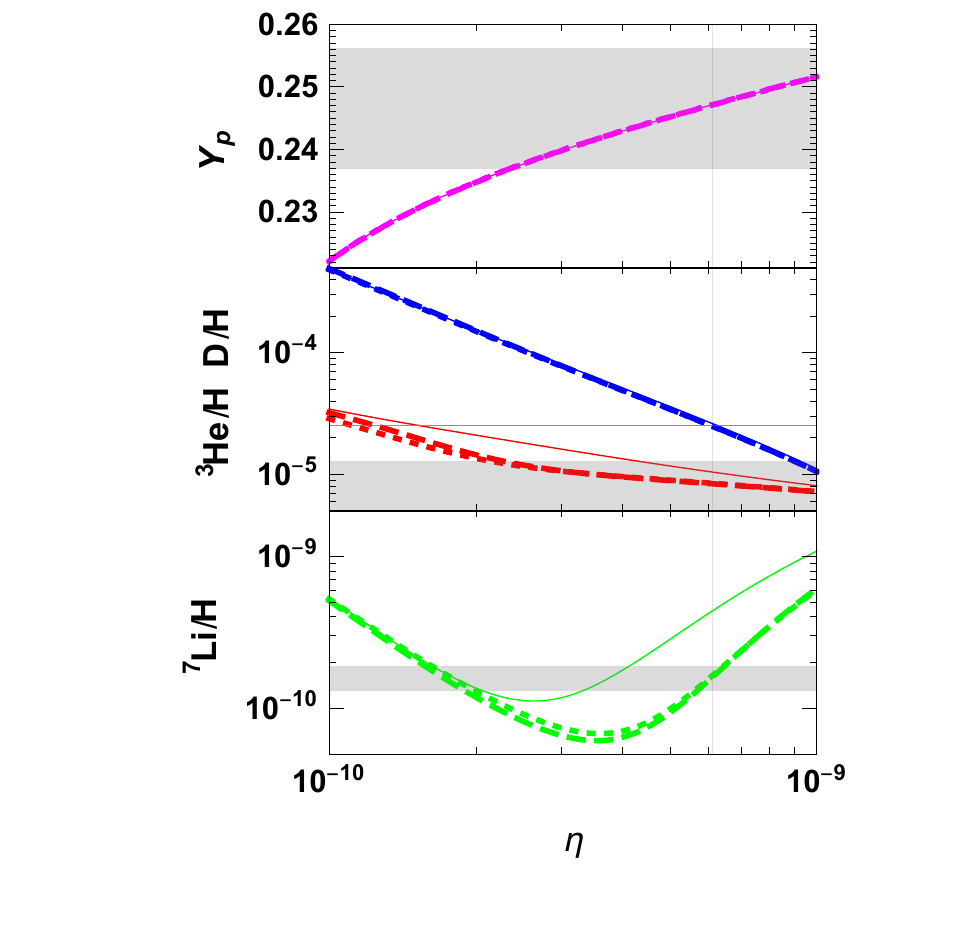}
\caption{The 
primodial abundances with respect to the
baryon-to-photon ratio $\eta$.
The solid lines are for the standard BBN,
and the dotted and dashed lines are for the parameter set
No. 1 and No. 4, respectively.
The vertical line at $\eta=6.10\times 10^{-10}$ denotes the 
value of the ratio $\eta$
obtained by the Planck~\cite{planck},
and the horizontal bands in gray 
are the observed light element abundances\cite{olive}.
}
\label{SchrammPlot}
\end{figure}

In Fig.~\ref{SchrammPlot}, we plotted 
the abundances of the light elements as functions of
the baryon-to-photon ration $\eta$.
The solid lines are for the standard BBN
without the NTD,
and the dotted and dashed lines are for the parameter sets No. 1 and No. 4,
respectively.
It shows that
the $\atom{He}{}{4}$ and $\atom{D}{}{}$ abundances
are little changed, while
the $\atom{Li}{}{7}$ abundance is
substantially reduced,
as required to resolve the
``lithium problem".
The abundance of $\atom{He}{}{3}$/H
is also found to be
noticeably reduced by including the NTD.
The parameter sets Nos. 1 and 4
give us 
$\atom{He}{}{3}/\mbox{H} = 0.85\times 10^{-5}$,\footnote{
What is meant by $\atom{He}{}{3}/\mbox{H}$ here is the sum of
$\atom{He}{}{3}$/H and T/H in Table 2.}
which is consistent with
the upper limit evaluated in Ref.~\cite{bania},
$\atom{He}{}{3}/\mbox{H} \le (1.1 \pm 0.2) \times 10^{-5}$.
It is to be noted that
the primordial abundance of $\atom{He}{}{3}$
is still 
uncertain,
the only data available coming from the Solar 
system and solar-matallicity HII regions 
in the Galaxy~\cite{bania}.
For this reason, we have not 
included the $\atom{He}{}{3}$ primordial abundance
in our chi-square estimation;
see Eq.(\ref{chi2-def}).

\section{Discussions}

We have studied the consequences of
introducing a small fraction of non-thermal particles   
during the BBN process,
allowing its magnitude 
to be time-dependent so that it contributes
only for a certain period.
This work may be regarded as an extension
of the work by Kang et al.\cite{kang}
where the magnitude was treated
as time-independent.
This extension, however, widens enormously  
the parametric space to be explored.
As the first step in this direction,
we have made
the assumption of Eq.~(\ref{assume}).
Therefore,
the contributions from the NTD of particles
have been
modeled 
in such a way that the average reaction rates 
are a superposition of two Maxwellian reaction rates 
of temperature $T$ and $T_{\rm NTD} = \zeta T$ given by Eq.~(\ref{assume}).
The calculations are based on the Kawano code,
and thus the advances in nuclear cross sections
made after the establishment of the Kawano code 
have not been taken into account.

With these caveats mentioned,
we are in the position to discuss what is found.
By scanning about half a million points in the parameter space,
we have found more than 130 points which have $\chi^2$ less than 1.
Among them, twelve points have $\chi^2<0.1$,
in good agreements with the observational data.
Those minima are found to be scattered around in the parameter space.

When the width $\Delta$ is small,
the parameters
with small values of $\chi^2$ turn out to be located in a narrow strip
in the parametric space around
$T_{09} \simeq 0.4$
and
a strong correlation between $\varepsilon_0$ and $\zeta$ is observed.
When the width parameter is as small as 0.0625,
the parameter sets with small $\chi^2$ values exist over a range
with $0.37 \lesssim T_{09} \lesssim 0.43$.
The corresponding temporal range is $1300 \gtrsim t \gtrsim 950$ seconds.
Our result is similar to the earlier study \cite{bailly}, where
it was discussed that
the stau-NLSP and gravitino-LSP system
with stau lifetime $\tau \simeq 10^3\ \mbox{s}$ 
could resolve the lithium problem with some representative values of
the model parameters.
If the width is as large as $\Delta=1$,
we could still find a large number of parameter sets with small $\chi^2$
which, however,
turns out to be scattered in rather a broad region
in the $T_{09}$ parametric space.
This may imply that there can be diverse NTD-induced mechanisms 
that can bring the BBN predictions to the observation data.
It is certainly necessary to refine the model
to overcome the above mentioned limitations
to identify the reaction channels
responsible for 
the cure of the lithium problem.

It would be very useful and interesting
to understand what happens 
when some of the underlying assumptions are released.
Furthermore, we have not yet discussed ``chemical spectrum" of 
cosmic rays, which will enlarge the parametric space
enormously.
These extensions are under progress
by making use of an updated version
for the Kawano code.

\section*{Acknowledgements}
We would like to thank Chung Yeol Ryu and Sang-In Bak 
for valuable discussions.
This work was supported by 
the National Research Foundation of Korea (NRF) funded
by the Ministry of Education, Science and Technology
(2017R1A2B4012758
and 2013M7A1A1075764).
TSP was also partly supported by the Institute for Basic Science (IBS-R031-D1).

\end{document}